\documentclass[12pt]{iopart}
\usepackage{graphicx}
\eqnobysec

\newcommand{\beq}{\begin{equation}}
\newcommand{\eeq}{\end{equation}}
\newcommand{\beqa}{\begin{eqnarray}}
\newcommand{\eeqa}{\end{eqnarray}}
\renewcommand{\a}{\alpha}
\newcommand{\abs}[1]{\vert#1\vert}
\newcommand{\ac}{{\rm AC}}
\renewcommand{\b}{\beta}
\renewcommand{\c}{{\rm c}}
\newcommand{\cas}{\noindent $\bullet$ {\hskip 5pt}}
\renewcommand{\d}{{\rm d}}
\newcommand{\ds}{\displaystyle}
\newcommand{\eps}{\varepsilon}
\newcommand{\euler}{{\bf C}}
\renewcommand{\etal}{{\it et~al}}

\newcommand{\g}{\gamma}
\newcommand{\half}{{\textstyle{1\over2}}}
\renewcommand{\i}{{\rm i}}
\renewcommand{\max}{{\rm max}}
\newcommand{\mean}[1]{\langle#1\rangle}
\newcommand{\mud}{{\mu_{\rm d}}}
\newcommand{\mus}{{\mu_{\rm s}}}
\newcommand{\ove}{{\rm over}}
\newcommand{\rms}{{\rm rms}}
\renewcommand{\ss}{\textstyle}
\newcommand{\st}{{\rm st}}
\newcommand{\var}{\mathop{\rm var}\nolimits}
\newcommand{\w}{\tilde}
\newcommand{\zero}{{(0)}}
\newcommand{\xid}{{\xi_{\rm d}}}
\newcommand{\xis}{{\xi_{\rm s}}}
\newcommand{\E}{{\rm E}}

\newcommand{\G}{{\rm G}}
\renewcommand{\H}{{\rm H}}

\newcommand{\Int}{\mathop{\rm Int}\nolimits}

\begin{document}

\title{Power-law forgetting in synapses with metaplasticity}

\author{A Mehta$^1$ and J M Luck$^2$}

\address{$^1$ S N Bose National Centre for Basic Sciences, Block JD,
Sector 3, Salt Lake, Calcutta 700098, India}

\address{$^2$ Institut de Physique Th\'eorique, IPhT, CEA Saclay
and URA 2306, CNRS, 91191~Gif-sur-Yvette cedex, France}

\begin{abstract}
The idea of using metaplastic synapses to incorporate the separate storage
of long- and short-term memories via an array of hidden states
was put forward in the cascade model of Fusi~\etal.
In this paper, we devise and investigate two models of a metaplastic synapse
based on these general principles.
The main difference between the two models lies in their available mechanisms of decay,
when a contrarian event occurs after the build-up of a long-term memory.
In one case, this leads to the conversion of the long-term memory to a short-term memory
of the opposite kind,
while in the other, a long-term memory of the opposite kind may be generated as a result.
Appropriately enough, the response of both models
to short-term events is not affected by this difference in architecture.
On the contrary,
the transient response of both models, after long-term memories
have been created by the passage of sustained signals,
is rather different.
The asymptotic behaviour of both models is, however,
characterised by power-law forgetting with the same universal exponent.
\end{abstract}

\eads{\mailto{anita@bose.res.in},\mailto{jean-marc.luck@cea.fr}}

\maketitle

\section{Introduction}
\label{intro}

Human memories are known to be fickle, but they are also capable of being
elephantine.
While research in this field is longstanding~\cite{ebb} in the
field of psychology, it is only relatively recently that it has been attacked
from an interdisciplinary perspective.
The seminal work of Amit
and collaborators~\cite{a1,a2,amitbook} on neural networks was in large part responsible
for opening up the field to physicists~\cite{dg};
much the same can be said about the work of Hopfield~\cite{hop}.
The optimisation of learning on complex neuronal networks
has been a field in itself; it has generally assumed that memories are stored
via the abrupt change that occurs in the synapses connecting
neurons, when they are exposed to a particular pattern.
This picture is premised on the notion of binary synapses (`synaptic switches'),
which are a natural approximation to synapses possessing a finite
set of discrete states.
There is some experimental evidence~\cite{A,B} in their support,
and they have also been extensively used in earlier mathematical models
(see e.g.~\cite{p1,p2,p3}).

The above mechanism of synaptic plasticity has, however,
been shown to be rather inefficient when synapses
change permanently~\cite{amfuss}.
Pure plasticity indeed does not provide a mechanism
for protecting some memories while leaving room for other,
newer, memories to come in,
hence the need for the mechanism of metaplasticity~\cite{a2}.
In order to improve performance,
Fusi~\etal~\cite{fusi} proposed a cascade model of a synapse with many hidden states,
which they claimed was able to store long-term memories more efficiently,
with a decay that was power-law rather than exponential in time.
Such power-law forgetting has in fact also been observed
experimentally~\cite{we1,we2} (albeit at a behavioural rather than a synaptic level).
This issue forms the focus of the current paper, where
we also put Fusi~\etal's cascade model on a more quantitative basis, by submitting
it to detailed questioning in a way that has not been done in either the
original work or in subsequent papers.
Another aim of our work is to see whether the introduction of architectural
differences might induce important differences in behaviour:
we accordingly devise a model which has a different mechanism
for the decay of long-term memories, compared to the one of Fusi~\etal,
and compare the two~models.

The plan of this paper is as follows.
In Section~\ref{models} we define both models to be investigated.
Model~I is an extension of the original cascade model by Fusi~\etal,
whereas Model~II has a different architecture.
Both models however share the common feature that all the transition probabilities
decay exponentially with the level depth of the hidden states.
Section~\ref{formalism} presents the formalism of Markov chains used in this work.
The default states of the two models are studied
in Section~\ref{default}.
This allows us to identify some useful parameters, which
include static and dynamical lengths $\xis$ and~$\xid$ relevant to the problem.
Section~\ref{ltpdc} is devoted to the response of both models
to a single long-term potentiating (LTP) input signal and to a DC signal
(sustained LTP signal); here we also provide an investigation of
universal asymptotic power-law forgetting (common to both models)
and of the non-universal transient forgetting specific to Model~II.
In Section~\ref{stn} we study the signal-to-noise ratio
which emerges from an investigation of fluctuations around the default state,
while in Section~\ref{variety} we illustrate the response of the models
to a selection of specific time-dependent input signals.
While some of these signals may be seen to be biologically unrealistic,
they are necessary for a systematic study of our models,
viewed from a physicist's perspective as signal processing units.
Finally, we discuss our results in Section~\ref{discussion}.
Appendix~A contains a detailed investigation of the problem of the logarithmic walker,
whereas Appendix~B examines the transient behaviour of both models,
and includes a derivation of the non-universal transient exponent of Model~II.

\section{The models}
\label{models}

In this section, we define the models to be studied
and introduce some of the ideas relevant to our investigations.
Synapses can respond differently to an incoming
action potential, in a way that could change with time~\cite{plast}: if a particular
stimulation paradigm leads to a persistent increase in response, this leads to
the long-term potentiation of synapses (LTP), whereas long-term depression
(LTD) corresponds to the opposite limit.
This change in the strength of a synapse
from a weak to a strong state and vice versa is referred to as {\it synaptic
plasticity} and forms the basis of the current understanding of learning and
memory, when applied to the many interconnected networks of synapses in the brain.
If synapses are highly plastic, memories are quickly stored: however,
high plasticity also means that more and more memories are stored, generating
enough noise so that earlier memories are soon irretrievable.
Clearly, this is
at variance with the fact that long-term memories are quite ubiquitous in human
experience; it was to resolve this paradox that Fusi~\etal~\cite{fusi} devised
the cascade model which is the motivation for the present paper.

The pathbreaking idea behind the work of Fusi~\etal\ was that the introduction
of `hidden states' for a synapse would enable the delinking of
memory lifetimes from instantaneous signal response: while maintaining quick learning,
it would also be able to allow slow forgetting.
In the original cascade model of~\cite{fusi}, this was implemented
by the storage of memories at different `levels': the relaxation times for
the memories increased as a function of depth.
It was assumed that
short-term memories, stored at the uppermost levels, would decay as a
consequence of their replacement by other short-term memories (`noise').
On the other hand, longer-lasting memories remained largely immune to such
noise as they were stored at the deeper levels, which were accessible only rarely.
This hierarchy of timescales models
the phenomenon of metaplasticity~\cite{meta1,meta2}.

In this work, we make a detailed comparison of two different models
of a metaplastic binary synapse with infinitely many hidden states (levels),
labelled by their depth $n=0,1,\dots$
At every discrete time step $t$, the synapse is subjected either to
an LTP signal (encoded as $\eps(t)=+1$) or to an LTD signal (encoded as $\eps(t)=-1$),
where $\eps(t)=\pm1$ is the instantaneous value
of the input signal at time $t$.

\begin{figure}[!ht]
\begin{center}
\includegraphics[angle=+00,width=.25\linewidth]{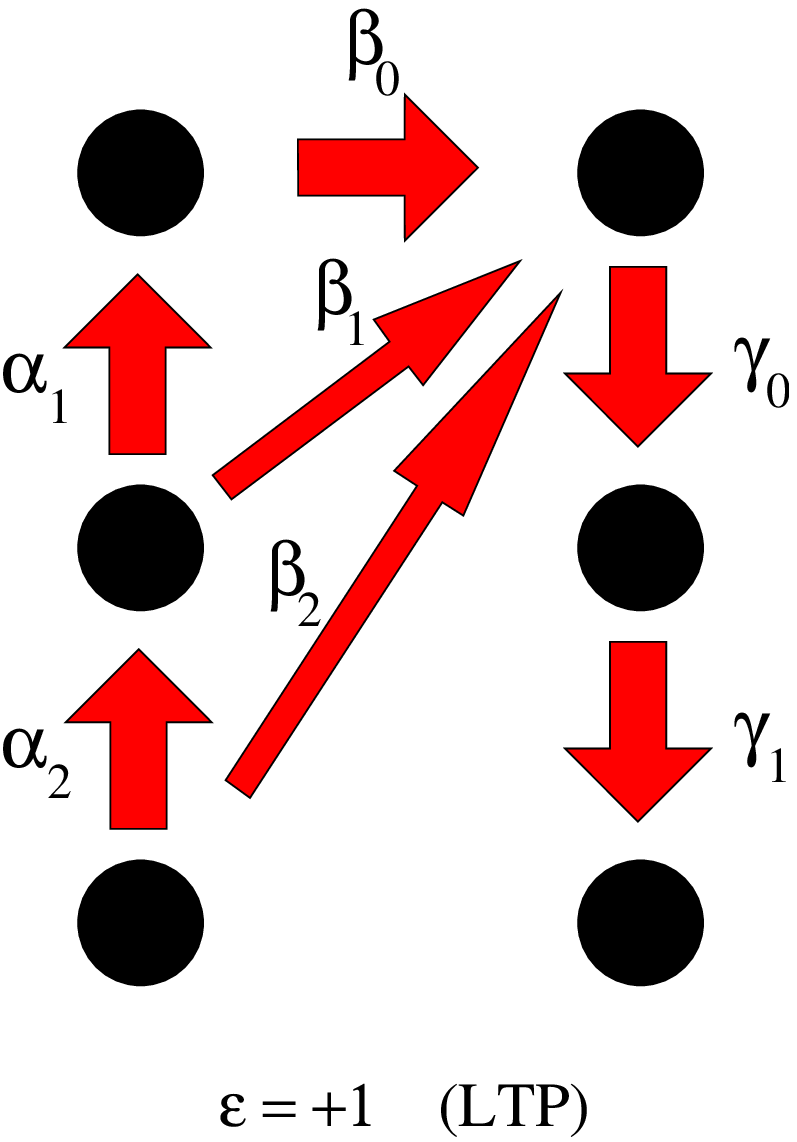}
{\hskip 25pt}
\includegraphics[angle=+00,width=.25\linewidth]{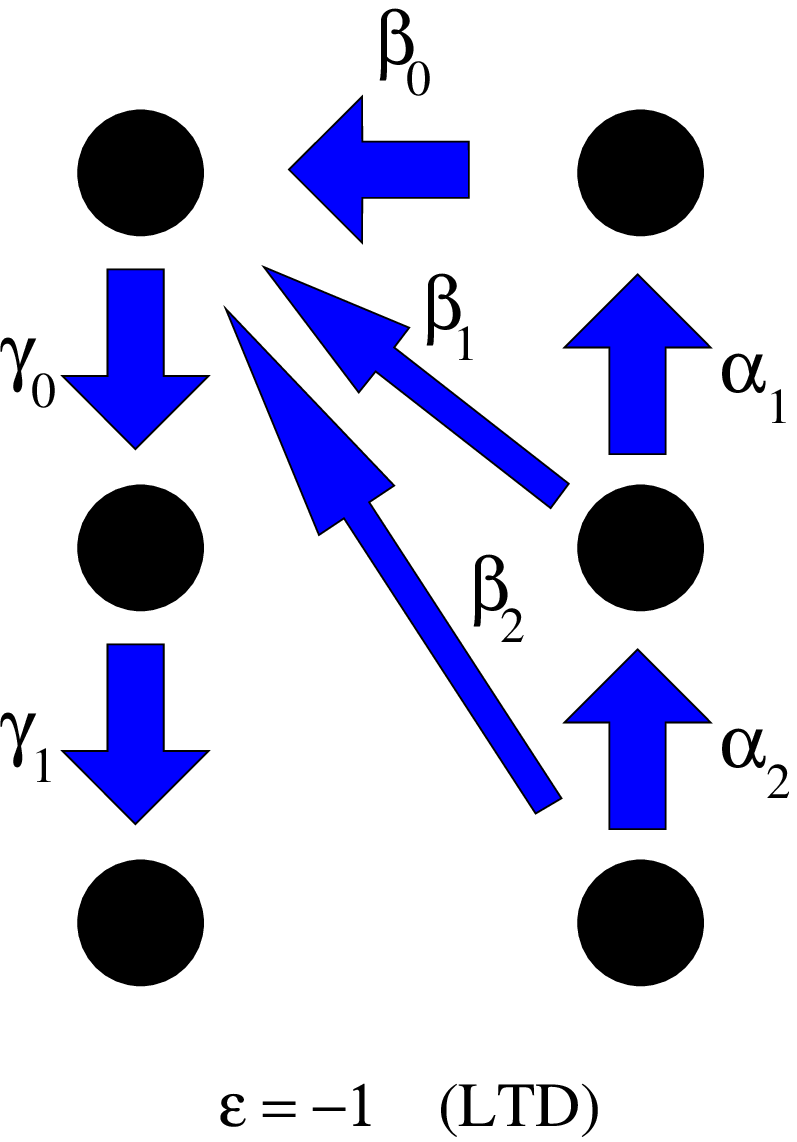}
\caption{\label{I}
Schematic representation of Model~I.
Arrows denote possible transitions in the presence of an LTP signal
($\eps=+1$, left panel)
and of an LTD signal
($\eps=-1$, right panel).
Corresponding transition probabilities are indicated.
In each panel, the left (resp.~right) column corresponds to the $-$ (resp.~$+$) state.
The model studied in this work is actually infinitely deep.}
\end{center}
\end{figure}

The first model (Model~I), defined in Figure~\ref{I},
is an extension of the original cascade model proposed by Fusi~\etal~\cite{fusi}.
The application of an LTP signal can have three effects:

\cas
If the synapse is in its $-$ state at depth $n$,
it may climb one level $(n\to n-1)$ with probability~$\a_n$.
(This move was absent in the original model.)

\cas
If it is in its $-$ state at depth $n$,
it may alternatively hop to the {\it uppermost} $+$ state with probability~$\b_n$.

\cas
If it is already in its $+$ state at depth $n$,
it may fall one level $(n\to n+1)$ with probability~$\g_n$.

Long-term memories will be stored in the deepest levels of the synapse,
because of the persistent application of characteristic signals.
The effect of noise on such a long-term memory is,
in the context of this model, to replace a long-term memory by a short-term
memory of the opposite kind.
If, for example, the signal is composed of entirely LTP events, an isolated
LTD event could be seen to represent the effect of noise.
In this case, the
Fusi model predicts that the signal is thrown from a deep positive level of
the synapse to the uppermost level of the negative pole.
Seen differently,
this mechanism converts a long-term memory of one kind to a short-term
memory of the opposite kind.

\begin{figure}[!ht]
\begin{center}
\includegraphics[angle=+00,width=.25\linewidth]{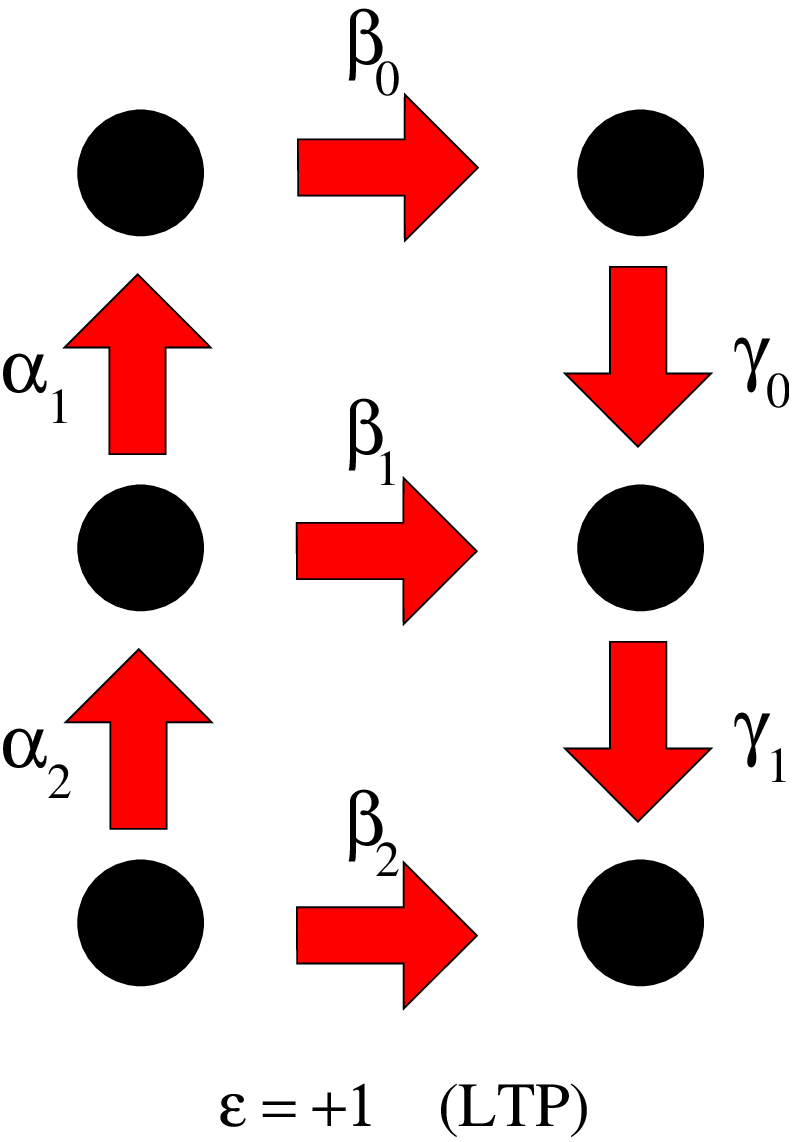}
{\hskip 25pt}
\includegraphics[angle=+00,width=.25\linewidth]{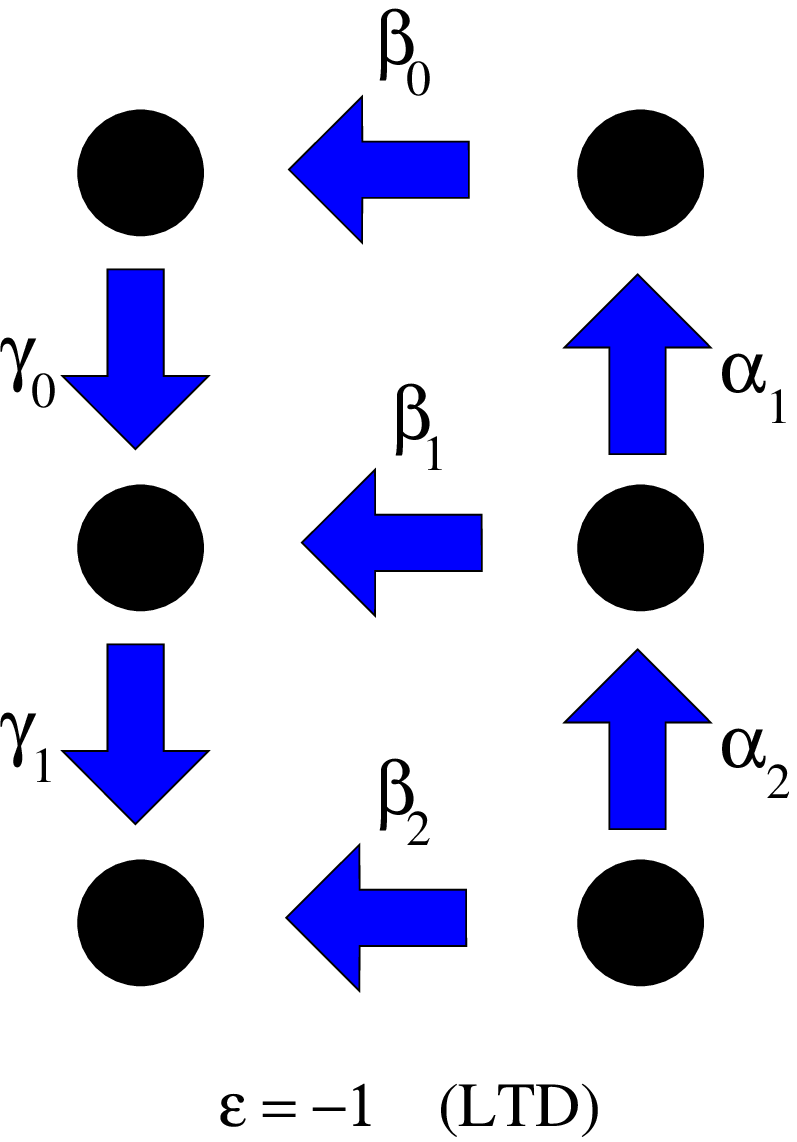}
\caption{\label{II}
Schematic representation of Model~II.
Same conventions as in Figure~\ref{I}.}
\end{center}
\end{figure}

It is however plausible that long-term memories of one kind could be replaced
by long-term memories of another kind
(e.g.~if a sudden event causes an abrupt change that is in its turn long-lasting).
Our Model~II, defined in Figure~\ref{II}, implements this mechanism.
The three outcomes of the application of an LTP signal are now as follows:

\cas
If the synapse is in its $-$ state at depth $n$,
it may climb one level $(n\to n-1)$ with probability~$\a_n$.

\cas
If it is in its $-$ state at depth $n$,
it may alternatively cross over to the $+$ state {\it at the same level} with
probability~$\b_n$.

\cas
If it is already in its $+$ state at depth $n$,
it may fall one level $(n\to n+1)$ with probability~$\g_n$.

Along the lines of Fusi~\etal~\cite{fusi},
the transition probabilities of both models
are assumed to decay exponentially with level depth $n$:
\beq
\a_n=\a\e^{-(n-1)\mud},\qquad
\b_n=\b\e^{-n\mud},\qquad
\g_n=\g\e^{-n\mud}.
\label{rates}
\eeq
The corresponding characteristic length,
\beq
\xid=\frac{1}{\mud},
\eeq
is one of the key ingredients of the models,
which measures the number of fast levels at the top of the synapse.
It will be referred to as the {\it dynamical length} of the problem.
The choice made in~\cite{fusi} corresponds to $\e^{-\mud}=\half$, i.e.,
$\mud=\ln 2$.
A different characteristic length, the static length $\xis$,
giving the number of occupied levels in the default state of the synapse,
will be introduced in Section~\ref{default}.

\section{Formalism}
\label{formalism}

We will make a detailed comparative analysis
of Model~I and Model~II, with a view to establishing similarities and
differences associated with their respective architectures.
In both cases the synapse is considered to be infinitely deep,
with levels numbered by $n=0,1,\dots$
We use the language of stochastic processes~\cite{kampen},
and in particular the formalism of inhomogeneous Markov chains.\footnote{This
formalism is a discrete-time analogue of that used extensively
in the mathematical literature,
to study e.g.~birth and death processes or queuing
processes~\cite{feller,karlin}.}

The basic quantities are the probabilities~$P_n(t)$ (resp.~$Q_n(t)$)
for the synapse to be in the $-$ state (resp.~in the $+$ state)
at level $n=0,1,\dots$ at time $t=0,1,\dots$
These probabilities can be combined in order to form quantities of interest:

\cas
Probability for the synapse to be in the $-$ state (resp.~in the $+$ state) at
time $t$, irrespective of level:
\beq
P(t)=\sum_{n\ge0}P_n(t),\qquad Q(t)=\sum_{n\ge0}Q_n(t)=1-P(t).
\eeq

\cas
Probability of being at level $n$ at time $t$, irrespective of state:
\beq
S_n(t)=P_n(t)+Q_n(t).
\eeq

\cas
Mean level depth
\beq
\mean{n(t)}=\sum_{n\ge0}nS_n(t).
\eeq

\cas
Level-resolved polarisation (output signal) of level $n$
and total polarisation of the synapse at time $t$:
\beq
D_n(t)=Q_n(t)-P_n(t),\qquad D(t)=\sum_{n\ge0}D_n(t)=Q(t)-P(t).
\label{ddef}
\eeq

We have the inequalities
\beq
\abs{D_n(t)}\le S_n(t),\qquad\abs{D(t)}\le1.
\eeq

The probabilities $P_n(t)$ and $Q_n(t)$ obey the following dynamical equations,
whose form is characteristic of Markov chains:

\cas
Model~I, $\eps(t+1)=+1$ (see Figure~\ref{I}, left):
\nopagebreak

\beq
\matrix{
P_n(t+1)=(1-\a_n-\b_n)P_n(t)+\a_{n+1}P_{n+1}(t),\hfill\cr
Q_n(t+1)=(1-\g_n)Q_n(t)+\g_{n-1}Q_{n-1}(t)+\delta_{n0}\w P(t),\hfill
}
\label{I+}
\eeq

\cas
Model~I, $\eps(t+1)=-1$ (see Figure~\ref{I}, right):
\nopagebreak

\beq
\matrix{
P_n(t+1)=(1-\g_n)P_n(t)+\g_{n-1}P_{n-1}(t)+\delta_{n0}\w Q(t),\hfill\cr
Q_n(t+1)=(1-\a_n-\b_n)Q_n(t)+\a_{n+1}Q_{n+1}(t),\hfill
}
\label{I-}
\eeq
with
\beq
\ds\w P(t)=\sum_{n\ge0}\b_nP_n(t),\qquad
\ds\w Q(t)=\sum_{n\ge0}\b_nQ_n(t).
\eeq

\cas
Model~II, $\eps(t+1)=+1$ (see Figure~\ref{II}, left):
\nopagebreak

\beq
\matrix{
P_n(t+1)=(1-\a_n-\b_n)P_n(t)+\a_{n+1}P_{n+1}(t),\hfill\cr
Q_n(t+1)=(1-\g_n)Q_n(t)+\g_{n-1}Q_{n-1}(t)+\b_nP_n(t).\hfill
}
\label{II+}
\eeq

\cas
Model~II, $\eps(t+1)=-1$ (see Figure~\ref{II}, right):
\nopagebreak

\beq
\matrix{
P_n(t+1)=(1-\g_n)P_n(t)+\g_{n-1}P_{n-1}(t)+\b_nQ_n(t),\hfill\cr
Q_n(t+1)=(1-\a_n-\b_n)Q_n(t)+\a_{n+1}Q_{n+1}(t).\hfill
}
\label{II-}
\eeq

\section{Default state and parameter space}
\label{default}

We here investigate the default state of the synapse,
which is the {\it average stationary state}
in the presence of a white-noise input signal.
White-noise input is defined by choosing at each time step
\beq
\eps(t)=\left\{\matrix{
+1\hfill&\hbox{with probability\ }\half,\cr
-1\hfill&\hbox{with probability\ }\half.
}\right.
\label{randef}
\eeq

In the presence of a random input $\eps(t)$,
the probabilities $P_n(t)$ and $Q_n(t)$ are themselves random.
We first evaluate the average response of the synapse,
encoded in the mean values of $P_n(t)$ and~$Q_n(t)$
with respect to the random input signal.
For simplicity, we continue to use the notation
$P_n(t)$ and $Q_n(t)$ for the {\it average} probabilities,
and $S_n(t)$ and $D_n(t)$ for their sums and differences.
As the values $\eps(t)$ of the input signal are independent of each other,
the equations obeyed by the mean probabilities
are the arithmetical means of~(\ref{I+}) and~(\ref{I-}) for Model~I,
and of~(\ref{II+}) and~(\ref{II-}) for Model~II.
The quantities $S_n(t)$ and $D_n(t)$ characterising the average response
therefore obey:

\cas
Model~I:
\nopagebreak

\beq
\matrix{
S_n(t+1)=S_n(t)+\half(\g_{n-1}S_{n-1}(t)+\a_{n+1}S_{n+1}(t))\hfill\cr
{\hskip 88.5pt}-\half(\a_n+\b_n+\g_n)S_n(t)+\half\delta_{n0}\w S(t),\hfill\cr
D_n(t+1)=D_n(t)+\half(\g_{n-1}D_{n-1}(t)+\a_{n+1}D_{n+1}(t))\hfill\cr
{\hskip 93.5pt}-\half(\a_n+\b_n+\g_n)D_n(t)-\half\delta_{n0}\w D(t),\hfill
}
\label{meanI}
\eeq
with
\beq
\ds\w S(t)=\sum_{n\ge0}\b_nS_n(t),\qquad
\ds\w D(t)=\sum_{n\ge0}\b_nD_n(t).
\eeq

\cas
Model~II:
\nopagebreak

\beq
\matrix{
S_n(t+1)=S_n(t)+\half(\g_{n-1}S_{n-1}(t)+\a_{n+1}S_{n+1}(t))\hfill\cr
{\hskip 88.5pt}-\half(\a_n+\g_n)S_n(t),\hfill\cr
D_n(t+1)=D_n(t)+\half(\g_{n-1}D_{n-1}(t)+\a_{n+1}D_{n+1}(t))\hfill\cr
{\hskip 93.5pt}-\half(\a_n+2\b_n+\g_n)D_n(t).\hfill
}
\label{meanII}
\eeq

The default state is characterised by
the time-independent solution to~(\ref{meanI}) or~(\ref{meanII}).
The latter is of the form
\beq
S_n^\st=(1-\e^{-\mus})\e^{-n\mus},\qquad D_n^\st=0,
\label{sdstat}
\eeq
i.e.,
\beq
P_n^\st=Q_n^\st=\half(1-\e^{-\mus})\e^{-n\mus}.
\label{pqstat}
\eeq

The default state is appropriately featureless.
It is unpolarised, as it should be for a symmetric synapse.
Furthermore, the occupation probabilities obey a simple exponen\-tial falloff
as a function of level depth.
The corresponding characteristic length,
\beq
\xis=\frac{1}{\mus},
\eeq
is referred to as the {\it static length} of the problem, and
gives a measure of the effective number of occupied levels in the default state.
The regime of most interest is where~$\xis$ is moderately large,
so that the default state extends over several levels.
The mean level depth
\beq
\mean{n}^\st=\frac{1}{\e^\mus-1}=\xis-\half+\cdots
\label{meanstat}
\eeq
is then essentially given by the static length.

The key role played by two characteristic lengths,
static ($\xis$) and dynamic ($\xid$),
is a striking similarity between this model and that of
a column of interacting grains investigated previously~\cite{us}.

In contrast to the dynamical length $\xid$, which is a free parameter,
the static length~$\xis$ is related to the values of the parameters $\a$, $\b$, and $\g$
in a model-dependent way.
Thus:

\cas
Model~I:
\nopagebreak

\beq
\g=\a\e^{-\mus}+\frac{\b}{\e^{\mus+\mud}-1}.
\label{abgI}
\eeq

\cas
Model~II:
\nopagebreak

\beq
\g=\a\e^{-\mus}.
\label{abgII}
\eeq

The above equations reveal the main difference between the two models
at the level of the default state.
The stationarity of the latter state involves balancing out
`upward' and `downward' moves arbitrarily deep within the system.
This goal is achieved in different ways in both models,
consistent with their structural differences.

In Model~II, a large static length $\xis$ is reached,
irrespective of $\b$,
when $\a$ and~$\g$ are nearly equal,
with a small bias in the upward direction:
\beq
\a-\g=(\e^\mus-1)\g\approx\frac{\g}{\xis}.
\eeq
The situation is very different for Model~I,
where non-local reinjection plays a key role.
The stationary profile of the response may become critical
(i.e., $\xis\to\infty$)
when a strong local downward bias is compensated
by strongly non-local upward moves:
\beq
\g-\a=\frac{\b}{\e^\mud-1}\approx\b\xid.
\eeq
This phenomenon is already at work in the original model by Fusi~\etal, where $\a=0$.

\begin{figure}[!ht]
\begin{center}
\includegraphics[angle=-90,width=.40\linewidth]{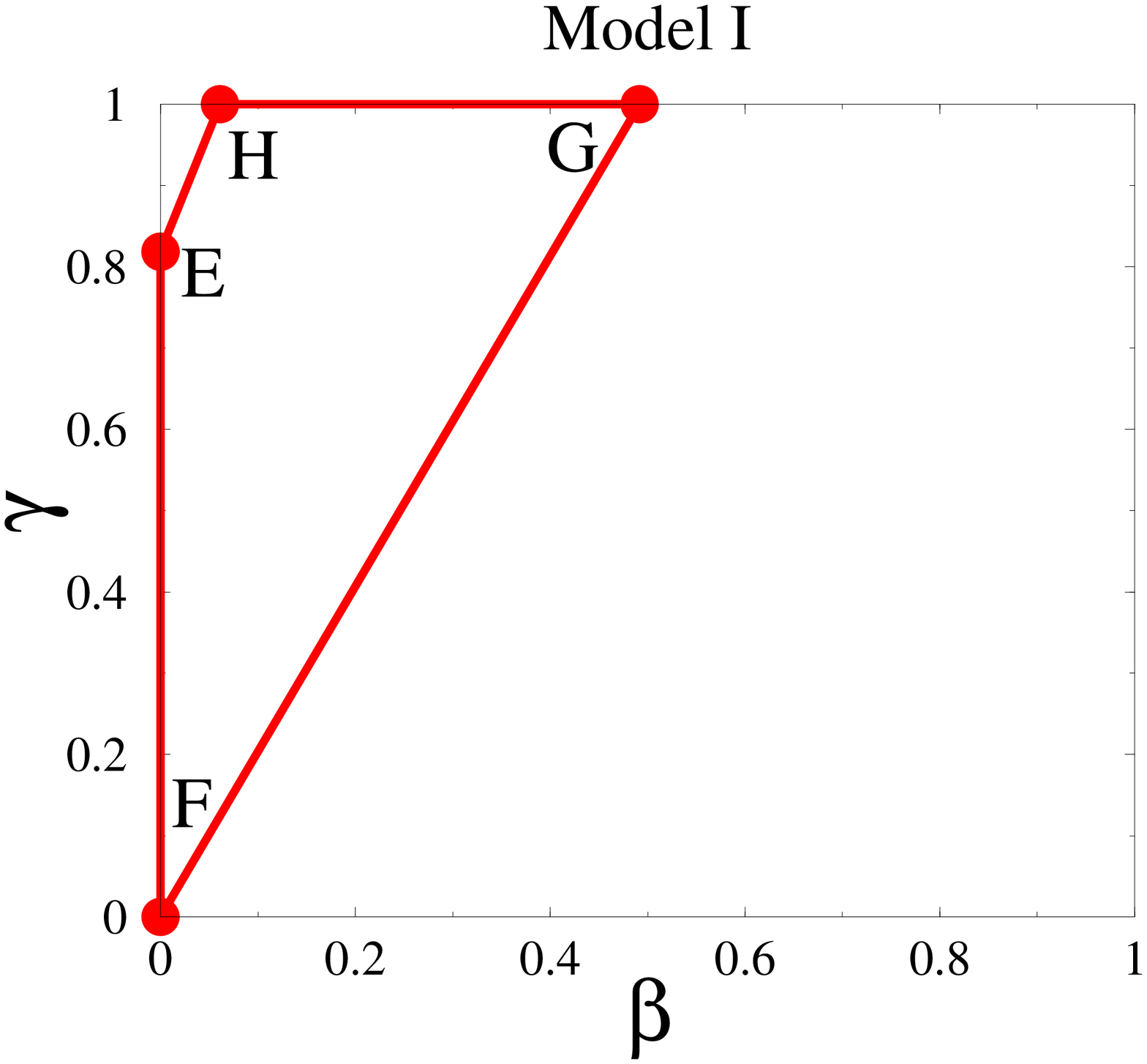}
{\hskip 25pt}
\includegraphics[angle=-90,width=.40\linewidth]{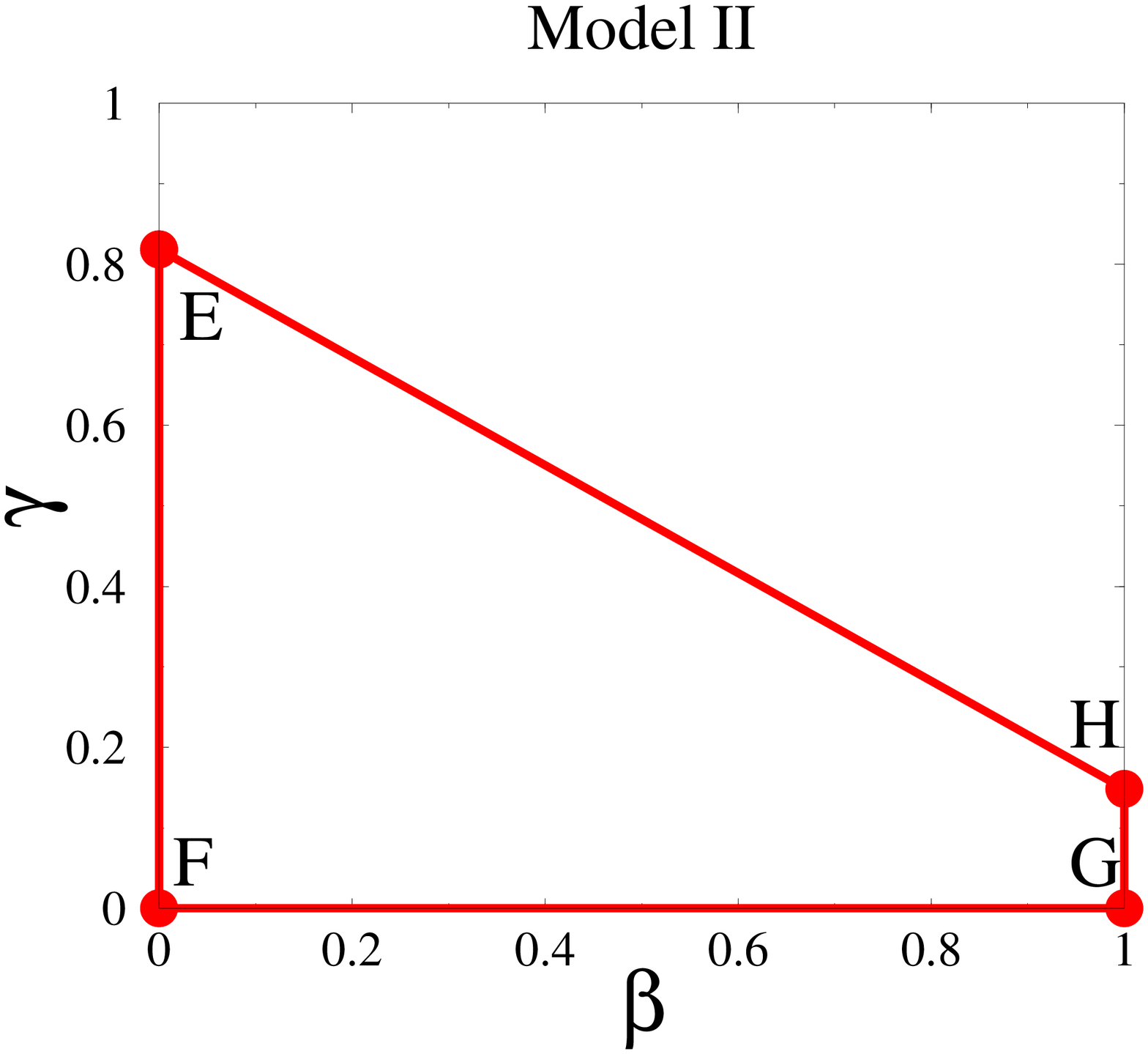}
\caption{\label{domains}
Domains of admissible values of $\b$ and $\g$ for both models with
$\xis=\xid=5$.}
\end{center}
\end{figure}

We now discuss the parameter space of both models.
The essential parameters are the static and dynamical lengths $\xis$ and $\xid$,
whose typical values are a few units.
For fixed $\xis$ and $\xid$,
$\a$, $\b$, and $\g$ are related by~(\ref{abgI}) or~(\ref{abgII}).
We choose to take $\b$ and $\g$ as our independent parameters.
Besides the condition that each of them is between 0 and~1,
they also fulfil (i) $\a\ge0$ and (ii) $\a_1+\b_1\le1$
(see~(\ref{I+}) or~(\ref{II+}) for $P_1(t+1)$).
For each model, the admissible values of $\b$ and $\g$
belong to a quadrangular domain EFGH,
shown in Figure~\ref{domains} for $\xis=\xid=5$.
In both cases, saturating condition (ii) yields the EH line.
The non-trivial coordinates of the vertices
as well as some special features, in the case of each model, are given below.

\cas
Model~I:
\nopagebreak

\beq
\g_\E=\e^{-\mus},\quad
\b_\G=\e^{\mus+\mud}-1,\quad
\b_\H=\e^\mud(\e^\mus-1)(\e^{\mus+\mud}-1).
\label{vertI}
\eeq
The maximal value of $\b$ for a fixed $\g$ lies on the FG line.
This
is the defining line for the original model of Fusi~\etal,
corresponding to the choice $\a=0$:

\beq
\b_\max(\g)=(\e^{\mus+\mud}-1)\g.
\label{bI}
\eeq

\cas
Model~II:
\nopagebreak

\beq
\g_\E=\e^{-\mus},\qquad
\g_\H=\e^{-\mus}(1-\e^{-\mud}).
\label{vertII}
\eeq
The maximal value of $\b$ for a fixed $\g$ lies on the (broken) EHG line:
\beq
\b_\max(\g)=\min(\e^\mud(1-\e^\mus\g),1).
\label{bII}
\eeq

The above expressions~(\ref{bI}) and~(\ref{bII})
for $\b_\max(\g)$ cross at the following critical value of $\g$:
\beq
\g_\c=\frac{\e^\mud}{2\e^{\mus+\mud}-1},
\eeq
so that Model~I has a smaller (resp.~larger) $\b_\max(\g)$
for $\g<\g_\c$ (resp.~$\g>\g_\c$).
This is a result to bear in mind,
as it turns out that the behaviour of many quantities of interest
is largely determined by $\b_\max(\g)$ (see e.g.~Figures~\ref{r} and~\ref{ac}).

Throughout the following, in numerical illustrations we use the parameter values
\beq
\xis=\xid=5,\qquad\hbox{(i.e.,}\quad\mus=\mud=0.2),\qquad\g=0.5,
\label{pars}
\eeq
unless otherwise stated.
For $\xis=\xid=5$, we have $\g_\c\approx0.615735$,
so that the chosen value of $\g$ is smaller than $\g_c$.
We have $\b_\max\approx0.245912$ for Model~I
and $\b_\max\approx0.475490$ for Model~II.

\section{Response to LTP input signals: power-law forgetting}
\label{ltpdc}

\subsection{Single LTP signal}

When a single LTP input signal is applied at time $t=1$ to the synapse
in its default state, it will get polarised in response, and thus `learn' the signal.
Later on, under the influence of a white-noise random input signal for times $t\ge2$,
it will forget the LTP signal, and return to its default state.
We will show that the process of forgetting is robust with respect to
the architectural differences between the two models, and is characterised by a
universal power law.

The polarised probability profile of the synapse at time $t=1$ is obtained by
acting once with equation~(\ref{I+}) or~(\ref{II+}) onto the default state~(\ref{pqstat}).
We thus obtain

\cas
Model~I:
\nopagebreak

\beq
\matrix{
P_0(1)=\half(1-\e^{-\mus})(1+\a\e^{-\mus}-\b),\hfill\cr
P_n(1)=\half(1-\e^{-\mus})\e^{-n\mus}\hfill\cr
{\hskip 29.5pt}+\half(1-\e^{-\mus})\e^{-n(\mus+\mud)}
(\a\e^{-\mus}-\a\e^{\mud}-\b)\quad\hfill&(n\ge1),\hfill\cr
Q_0(1)=\half(1-\e^{-\mus})(1+\b/(1-\e^{-\mus-\mud})-\g),\hfill\cr
Q_n(1)=\half(1-\e^{-\mus})\e^{-n\mus}\hfill\cr
{\hskip 31.5pt}+\half(1-\e^{-\mus})\e^{-n(\mus+\mud)}
(\e^{\mus+\mud}-1)\g\hfill&(n\ge1).\hfill
}
\label{Ione}
\eeq

\cas
Model~II:
\nopagebreak

\beq
\matrix{
P_0(1)=\half(1-\e^{-\mus})(1+\a\e^{-\mus}-\b),\hfill\cr
P_n(1)=\half(1-\e^{-\mus})\e^{-n\mus}\hfill\cr
{\hskip 29.5pt}+\half(1-\e^{-\mus})\e^{-n(\mus+\mud)}
(\a\e^{-\mus}-\a\e^{\mud}-\b)\quad\hfill&(n\ge1),\hfill\cr
Q_0(1)=\half(1-\e^{-\mus})(1+\b-\g),\hfill\cr
Q_n(1)=\half(1-\e^{-\mus})\e^{-n\mus}\hfill\cr
{\hskip 31.5pt}+\half(1-\e^{-\mus})\e^{-n(\mus+\mud)}
(\e^{\mus+\mud}-1)\g\hfill&(n\ge1).\hfill
}
\label{IIone}
\eeq

The instantaneous output signal,
i.e., the total polarisation $D(1)$ of the synapse just after the LTP signal,
takes the same value proportional to $\b$ for both models:
\beq
D(1)=\lambda_1\,\b,\qquad\lambda_1=\frac{1-\e^{-\mus}}{1-\e^{-\mus-\mud}}.
\label{done}
\eeq
For $\xis=\xid=5$ we have $\lambda_1\approx0.549833$.

\begin{figure}[!ht]
\begin{center}
\includegraphics[angle=-90,width=.45\linewidth]{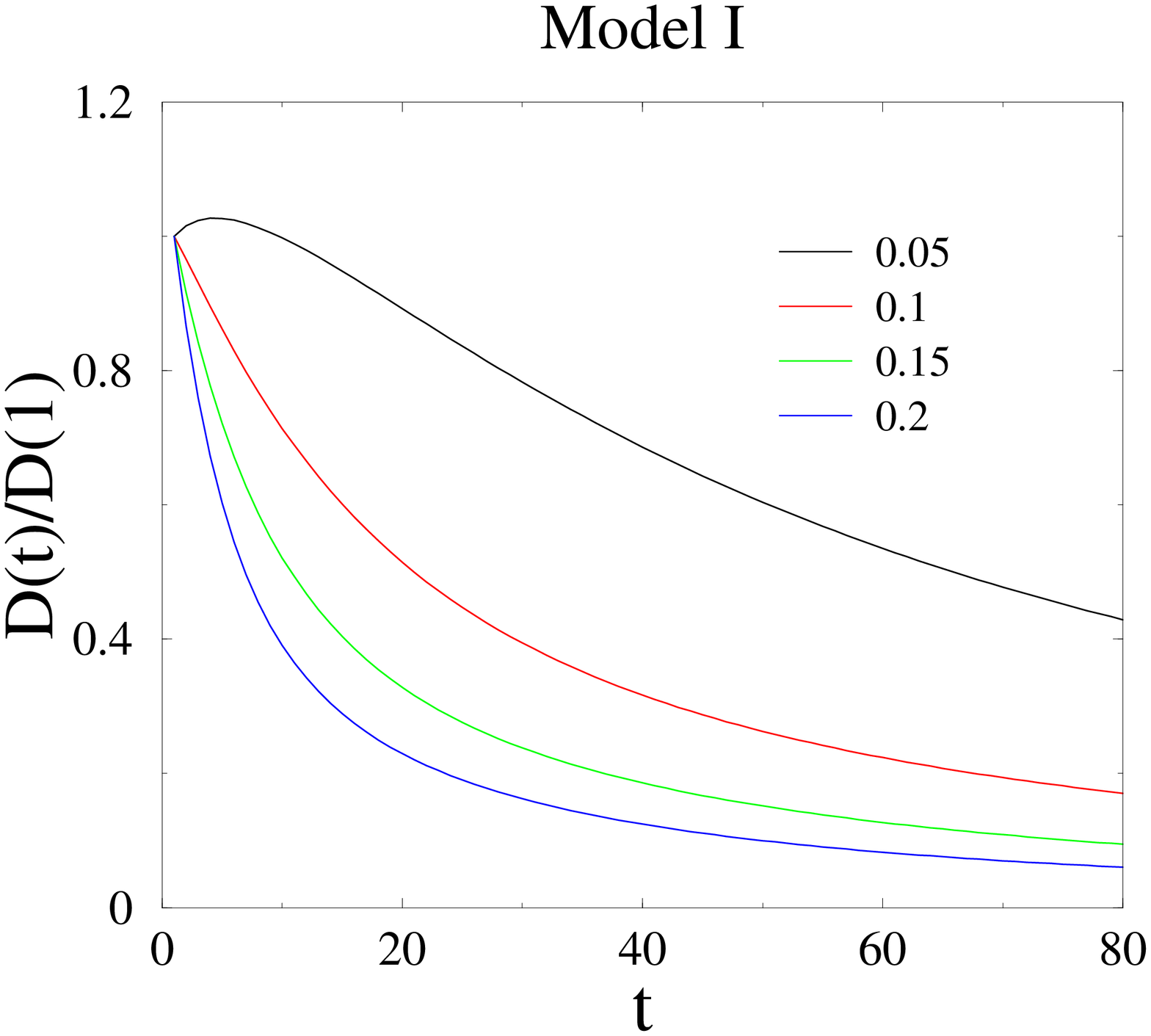}
{\hskip 10pt}
\includegraphics[angle=-90,width=.45\linewidth]{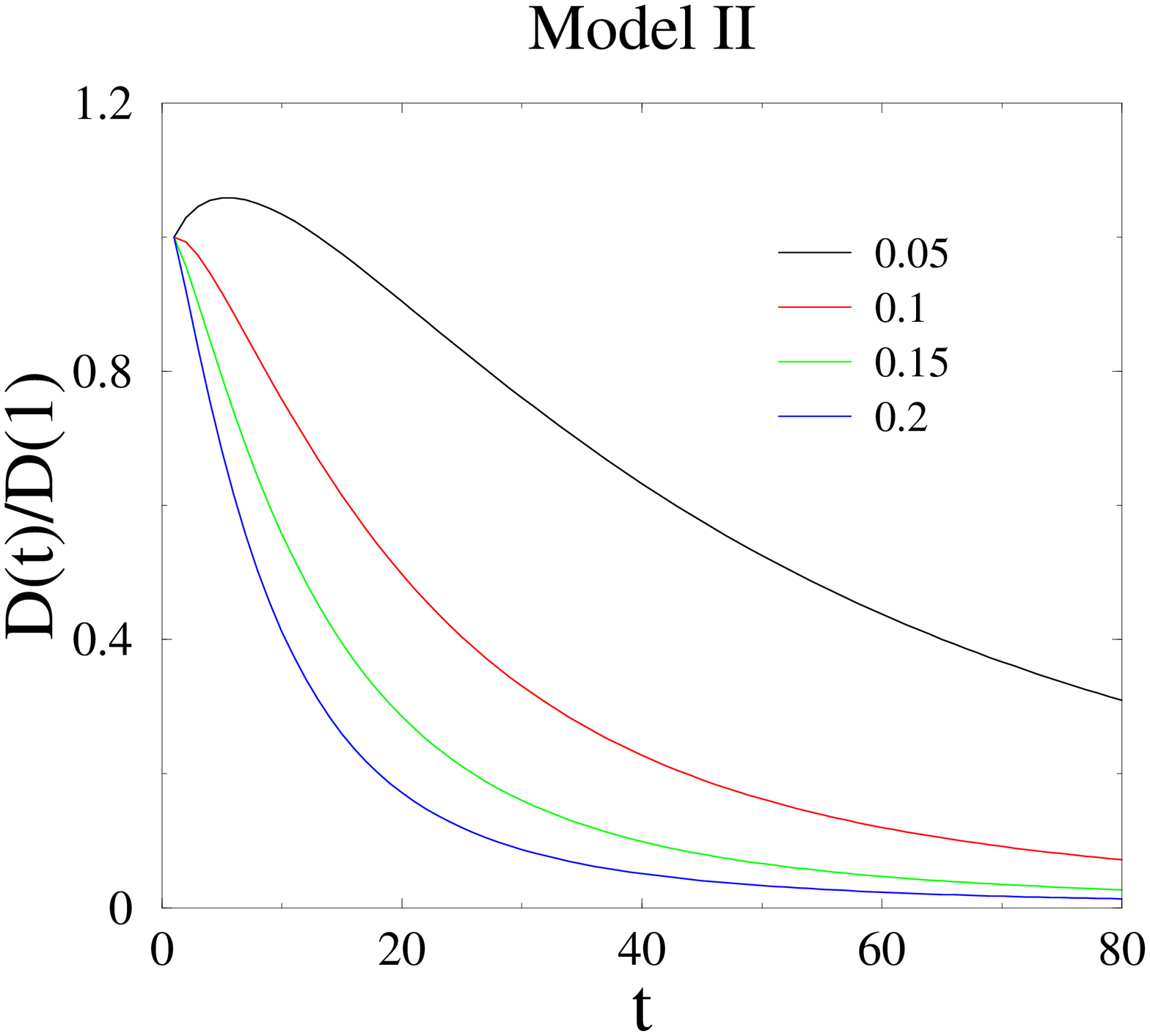}
\caption{\label{ltpred}
Plot of the reduced total polarisation $D(t)/D(1)$ after a single LTP signal,
against time $t$, for both models and several $\b$ (see legends).}
\end{center}
\end{figure}

The synapse then evolves under the influence of a white-noise random input
during the subsequent forgetting phase.
This evolution is described for Model~I by the action of the
recursion~(\ref{meanI})
on the probabilities~(\ref{Ione}),
and for Model~II by the action of~(\ref{meanII}) on~(\ref{IIone}).
Figure~\ref{ltpred} shows plots of the reduced polarisation signals $D(t)/D(1)$
against time $t$ for Model~I (left) and Model~II (right),
for several values of $\b$.
For small enough~$\b$, the polarisation overshoots,
i.e., it keeps increasing beyond $D(1)$
in a transient regime at the beginning of the forgetting phase.
The duration of this transient overshoot gets larger for smaller $\b$,
and formally diverges in the $\b\to0$ limit.

This paradoxical behaviour can be explained as follows.
In the forgetting phase, the total polarisation obeys the balance equation
\beq
D(t+1)-D(t)=-\sum_{n\ge0}\b_nD_n(t).
\label{balance}
\eeq
Generically, then, $D(t)$ decays to zero, as expected.
It may however grow in a transient regime,
leading to the overshoot mentioned above,
provided the initial polarisation profile is inhomogeneous enough
so as to satisfy both
\beq
D(t)=\sum_{n\ge0}D_n(t)>0\quad\hbox{and}\quad\sum_{n\ge0}\b_nD_n(t)<0.
\label{ineqs}
\eeq
For a single LTP signal,
the initial profile at time $t=1$
is such that $D_0(1)<0$, whereas $D_n(1)>0$ for $n\ge1$,
for both models and with $\b$ small (see~(\ref{Ione}),~(\ref{IIone})).
Since the rates~$\b_n$ fall off exponentially in $n$,
the initial profile is thus likely to obey the inequalities~(\ref{ineqs}),
thus leading to the overshoot.
In fact, it can be shown that the overshoot always occurs for
$\b<\b_\ove(\g)$, where:

\cas
Model~I:
\nopagebreak
\beq
\b_\ove(\g)=\frac{(1-\e^{-\mud})(1-\e^{-\mus-\mud})}{1-\e^{-\mus-2\mud}}\g.
\eeq

\cas
Model~II:
\nopagebreak
\beq
\b_\ove(\g)=(1-\e^{-\mud})\g.
\eeq
For the parameters~(\ref{pars})
we have $\b_\ove\approx0.066226$ for Model~I
and $\b_\ove\approx0.090634$ for Model~II.

To summarise, the instantaneous response $D(1)$ to an LTP signal
is proportional to~$\b$,
and therefore larger for larger $\b$; its subsequent decay is, however, fast
for large~$\b$ -- an undesirable feature --
whereas it is slow and even non-monotonic for smaller $\b$.
This suggests the absence of a natural criterion for defining an optimal $\b$,
where quick learning and slow forgetting might simultaneously occur at the synapse.

\subsection{Universal power-law forgetting}
\label{plf}

\begin{figure}[!ht]
\begin{center}
\includegraphics[angle=-90,width=.45\linewidth]{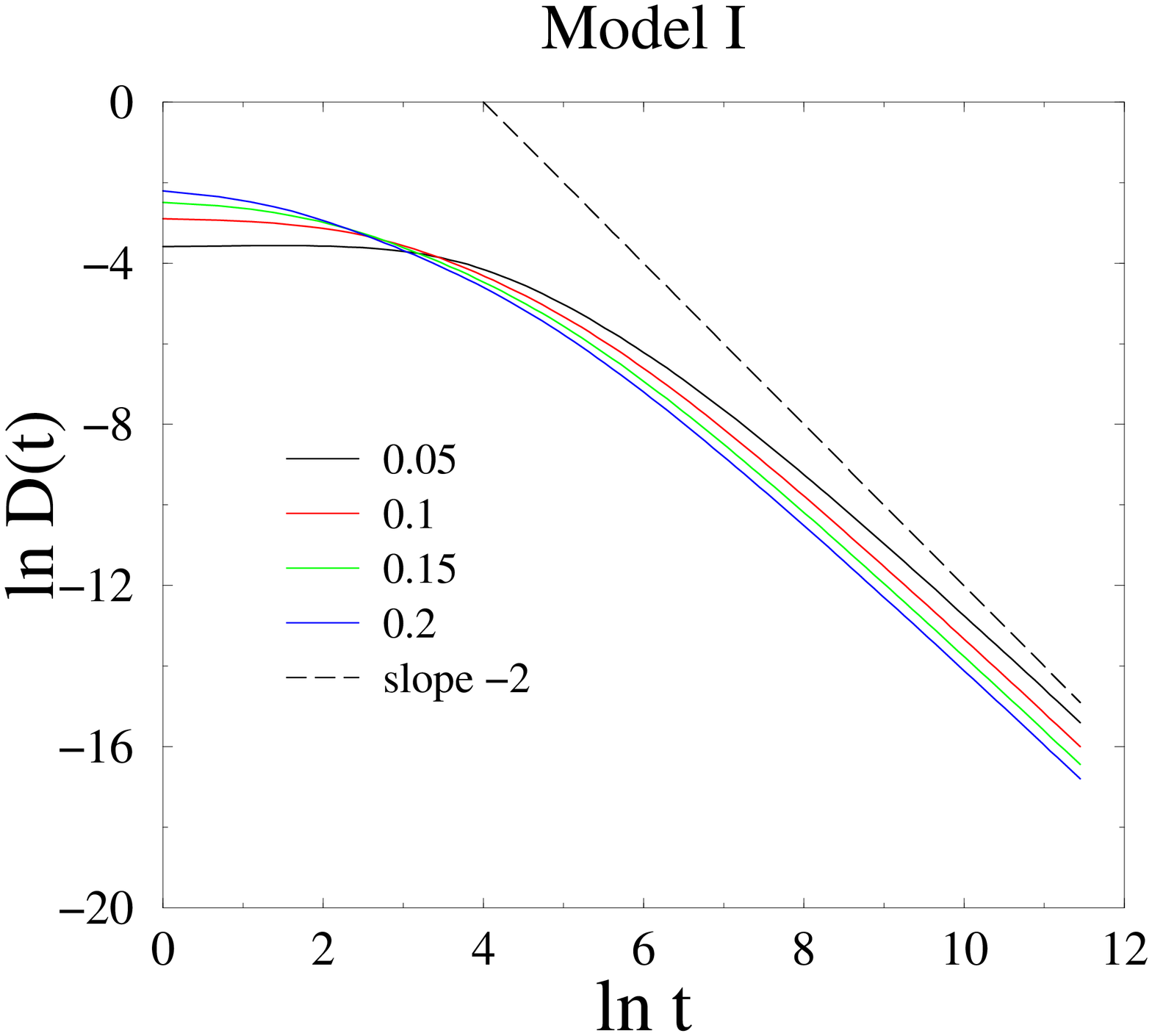}
{\hskip 10pt}
\includegraphics[angle=-90,width=.45\linewidth]{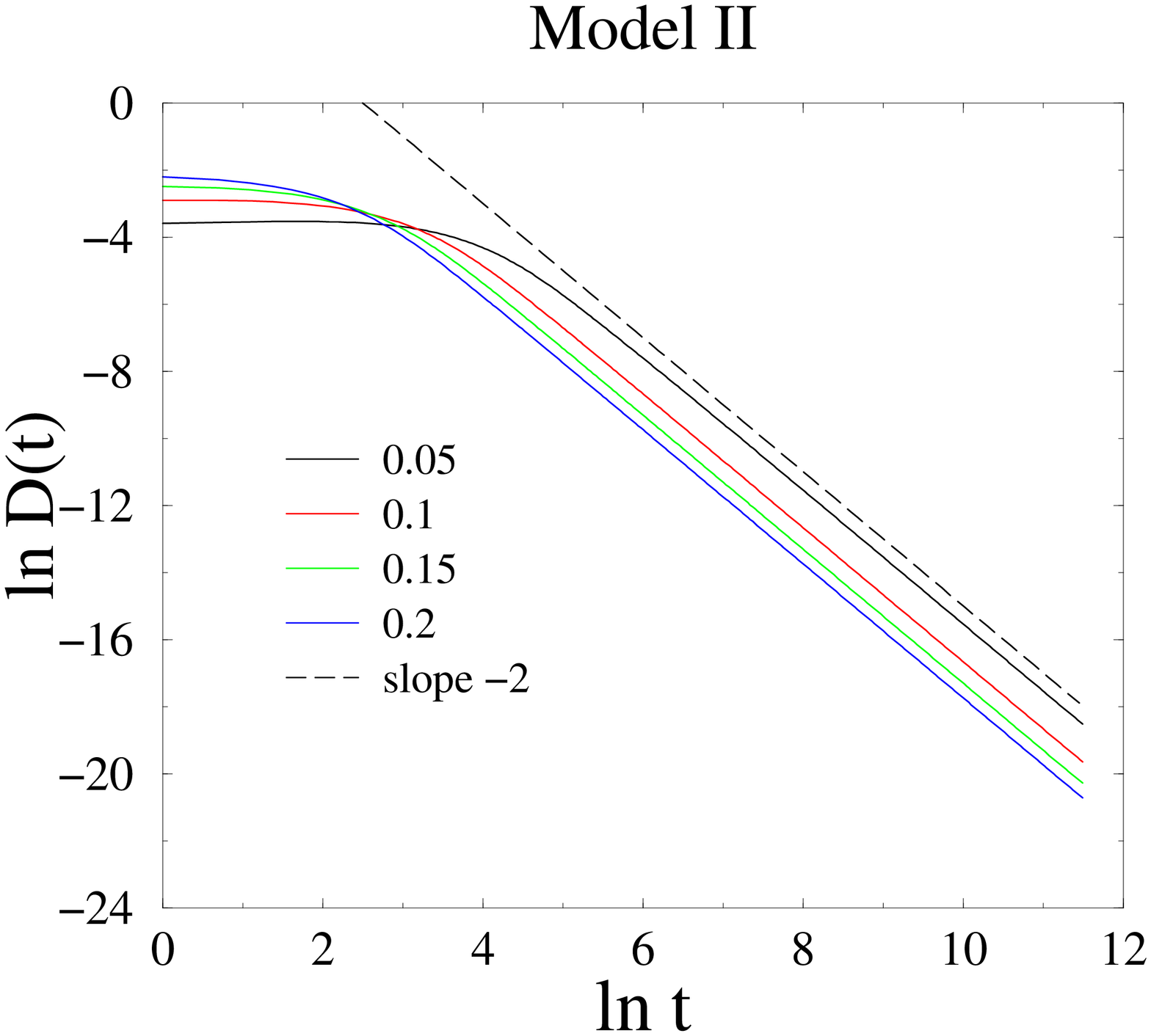}
\caption{\label{ltplog}
Log-log plot of the total polarisation $D(t)$
after a single LTP signal, against time~$t$,
for both models and several $\b$ (see legends).
The absolute slope of the dashed lines is the theoretical
value~(\ref{theta}), i.e., $\theta=2$.}
\end{center}
\end{figure}

The asymptotic fall-off of the total polarisation of the synapse
in response to a single LTP signal is illustrated in Figure~\ref{ltplog},
showing a log-log plot of $D(t)$ for much longer times (up to $t=10^5$).
The data for both our models show a common power-law decay:
thus, for our choice of parameter values, $D(t)\sim 1/t^2$
in both cases.\footnote{Corrections to the asymptotic power law are,
however, stronger for Model~I.}
This is known as {\it power-law forgetting}, which will be analysed below.

The expressions~(\ref{Ione}),~(\ref{IIone}) show
that the initial polarisation profile decays exponentially as a function
of level depth~$n$,~as
\beq
D_n(1)\sim\e^{-n(\mus+\mud)}.
\label{dinit}
\eeq
This exponential decay is governed by the product of the probabilities
$S_n^\st\sim\e^{-n\mus}$ in the default state (see~(\ref{pqstat}))
and the polarising rate $\b_n\sim\e^{-n\mud}$ (see~(\ref{rates})).

Now consider the synapse at a late stage of the forgetting phase ($t\gg1$).
The white-noise input essentially erases the polarisation profile
down to a level depth $n_*$ such that $\b_{n_*}t\sim1$.
More details on this derivation can be found in Appendix~A.
This~gives:
\beq
n_*\approx\xid\ln t.
\label{nstar}
\eeq
Of course, the only part of the polarisation that survives
at large times $t$ is the part which has not yet been forgotten:
this lives in the deeper levels ($n>n_*$),
where white noise has not yet erased the remnants of the memory.
The total polarisation is therefore expected to scale as $D_{n_*}(1)$.
Using the estimates~(\ref{dinit}) and~(\ref{nstar}),
we obtain an asymptotic power-law decay of the polarisation signal:
\beq
D(t)\sim t^{-\theta},
\label{dtheta}
\eeq
with
\beq
\theta=1+\frac{\mus}{\mud}=1+\frac{\xid}{\xis}.
\label{theta}
\eeq

The forgetting exponent $\theta$ thus obtained
only depends on the ratio of the static and dynamical lengths $\xis$ and $\xid$.
Its expression~(\ref{theta}) is universal,
in the sense that it holds irrespective of the model architecture,
and of the rates $\a$, $\b$, and $\g$,
besides the fact that~$\xis$ is related to the latter
parameters in a model-dependent way
(see~(\ref{abgI}),~(\ref{abgII})).
We would thus expect power-law forgetting with exponent $\theta$
to be manifested for a large class of learnt signals.

It is worth remarking here that,
if the synapse were finite rather than infinite, and consist of $N$ levels,
the power-law decay~(\ref{dtheta}) would be exponentially cutoff
at a time~$\tau$ such that $\b_N\tau\sim1$.
The cutoff timescale thus obtained,
\beq
\tau\sim\e^{N/\xid},
\eeq
is exponentially large in the ratio
of the number $N$ of levels to the dynamical length~$\xid$.

\subsection{DC signal (sustained LTP signal)}
\label{secdc}

We now turn to the investigation of a DC input signal,
i.e., a sustained LTP input signal
lasting for $T$ time steps ($\eps(t)=+1$ for $1\le t\le T$).
The synapse is again assumed to be initially in its default state.

\begin{figure}[!ht]
\begin{center}
\includegraphics[angle=-90,width=.45\linewidth]{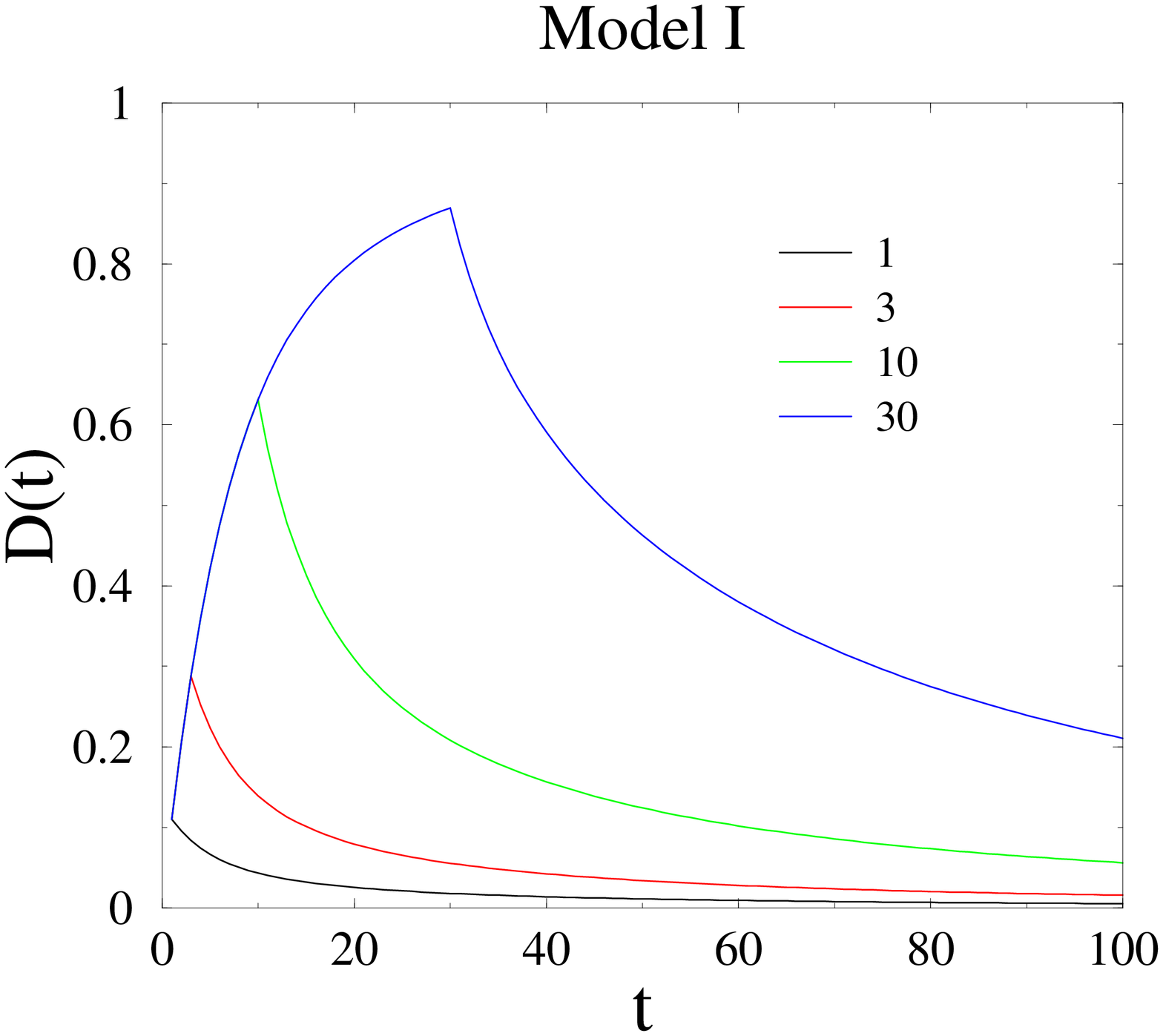}
{\hskip 10pt}
\includegraphics[angle=-90,width=.45\linewidth]{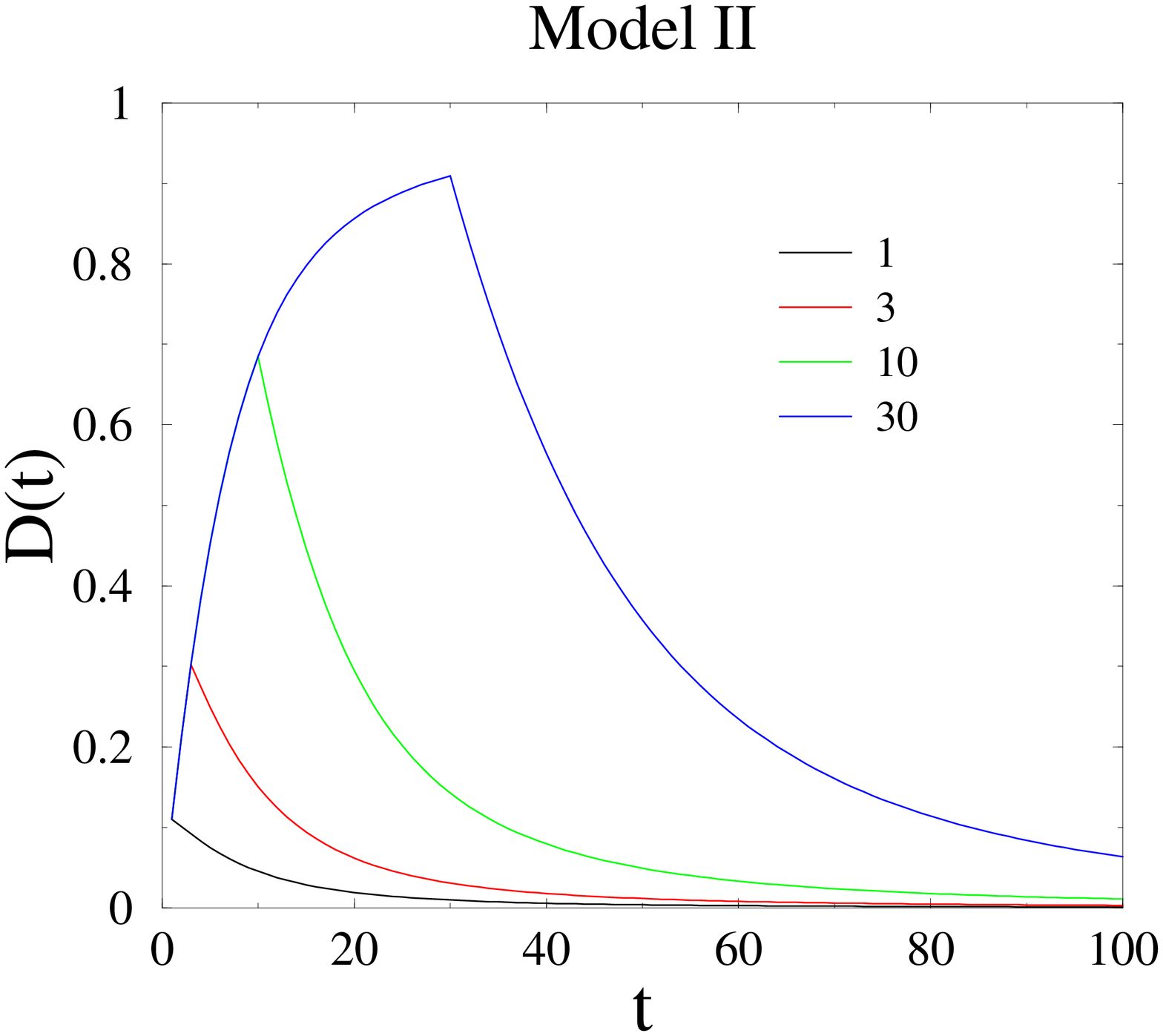}
\caption{\label{sted}
Plot of the total polarisation $D(t)$ of both models with $\b=0.2$,
against time $t$,
for several durations $T$ of the DC signal (see legends).}
\end{center}
\end{figure}

The learning and forgetting processes will be qualitatively similar to the above,
while novel qualitative features emerge deep in the DC regime,
i.e., when the duration of the LTP signal is long enough
so that the product $\b T$ is large.
In this regime, the synapse gets almost totally polarised
under the persistent action of the input signal.
This saturation phenomenon is illustrated in Figure~\ref{sted},
which shows the total polarisation~$D(t)$
of both models for several durations $T$ of the DC signal.

\begin{figure}[!ht]
\begin{center}
\includegraphics[angle=-90,width=.45\linewidth]{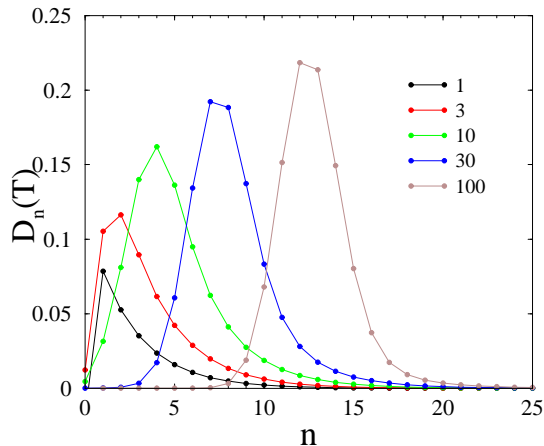}
\caption{\label{stef}
Plot of the polarisation profile $D_n(T)$
of Model~II with $\b=0.2$
at the end of the learning phase,
against level depth $n$,
for several durations~$T$ of the DC signal (see legend).}
\end{center}
\end{figure}

The synapse slowly builds up a long-term memory in the presence
of a long DC signal, as the polarisation profile moves to deeper and deeper levels.
This feature is illustrated in Figure~\ref{stef},
which shows a plot of the full polarisation profile of Model~II
at the end of the learning phase,
for several durations~$T$ of the DC signal.
When the synapse becomes fully polarised
in the late-time regime ($\b t\gg1$), the level polarisations become
approximately $D_n(t)=Q_n(t)$; for both models, the signals travel down the
synapse with exponentially decaying rates.
Thus, both~(\ref{I+}) and~(\ref{II+}) become:
\beq
D_n(t+1)=(1-\g_n)D_n(t)+\g_{n-1}D_{n-1}(t).
\eeq
The polarisation dynamics are therefore modelled
by that of the logarithmic walker (Appendix~A).
Thus, at the end of the learning phase ($t=T$),
the polarisation profile will have the form
of a sharply peaked traveling wave (see~(\ref{appq})),
around a mean depth which grows according to the logarithmic law
(see~(\ref{log}))
\beq
\mean{n}\approx\xid\ln\g T.
\label{nT}
\eeq

\begin{figure}[!ht]
\begin{center}
\includegraphics[angle=-90,width=.45\linewidth]{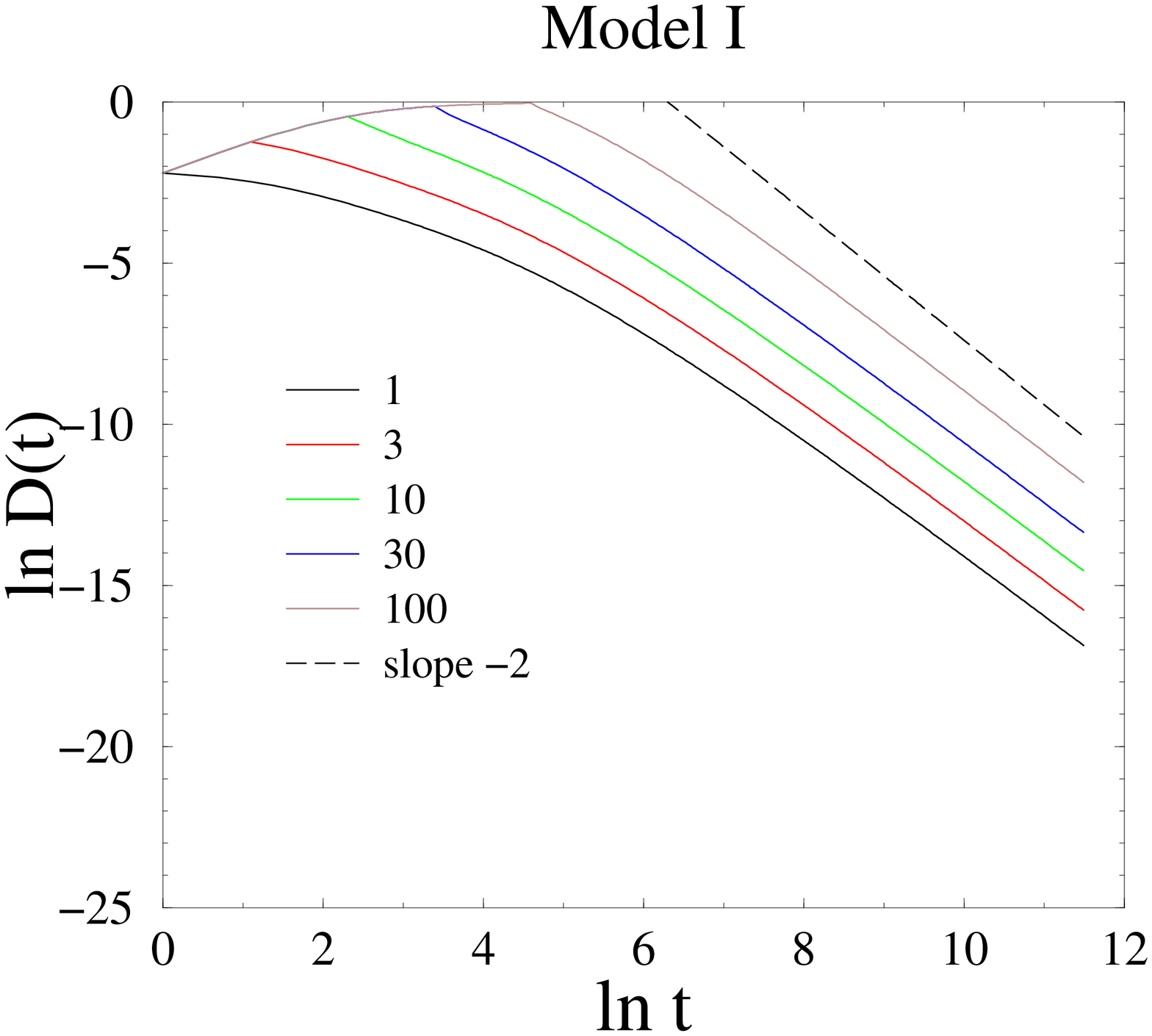}
{\hskip 10pt}
\includegraphics[angle=-90,width=.45\linewidth]{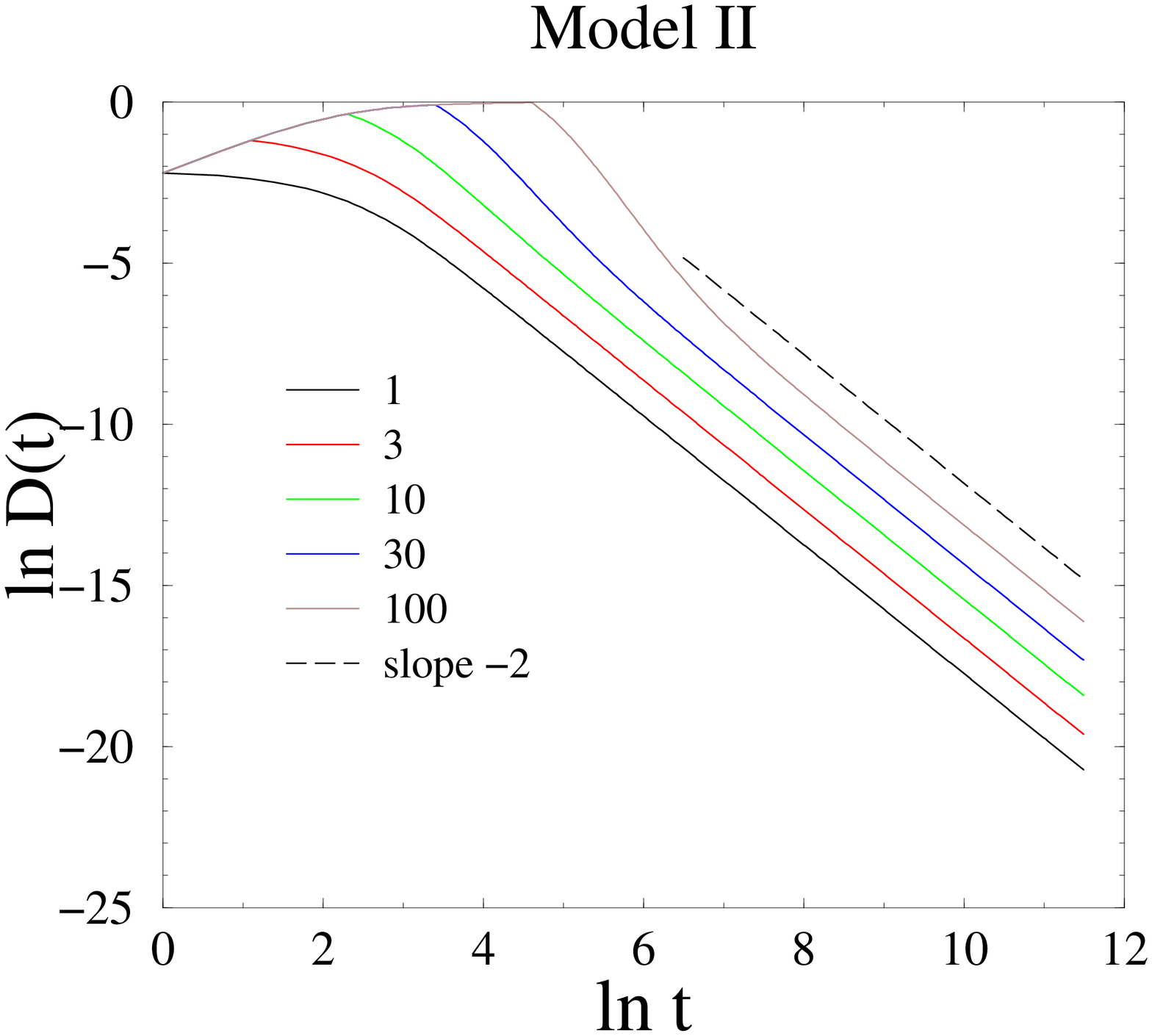}
\caption{\label{stedl}
Log-log plot of the total polarisation $D(t)$ against time~$t$,
for both models with $\b=0.2$ and several durations $T$ of the DC signal (see legends).
The absolute slope of the dashed lines is $\theta=2$.}
\end{center}
\end{figure}

We now turn to the decay of the total polarisation $D(t)$
generated by a sustained LTP signal, deep in the DC regime.
Figure~\ref{stedl} shows a log-log plot of $D(t)$
for several durations $T$ of the DC signal, and much longer observation times.
The polarisation decays via
the universal power law~(\ref{dtheta}),
irrespective of the length of the learning phase, driving home the universality
of power-law forgetting.

\subsection{Non-universal transient power-law forgetting}
\label{tplf}

So far, we have shown that most features of our two models of the synapse
are pretty robust to their different architectures: however, in the following,
we show an important phenomenon where the two models differ strongly,
for a sustained or persistent signal and at larger timescales.

\begin{figure}[!ht]
\begin{center}
\includegraphics[angle=-90,width=.45\linewidth]{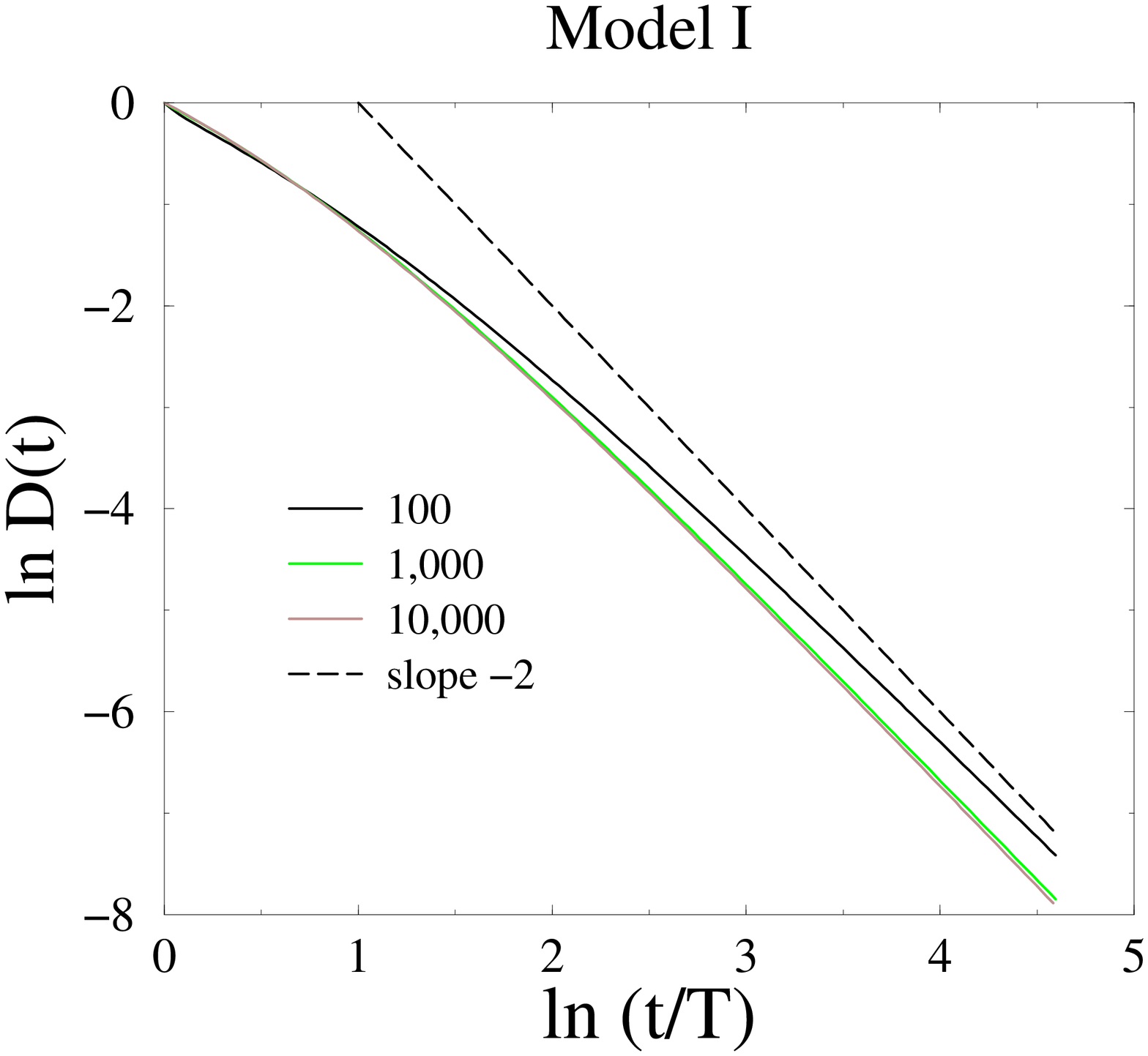}
{\hskip 10pt}
\includegraphics[angle=-90,width=.45\linewidth]{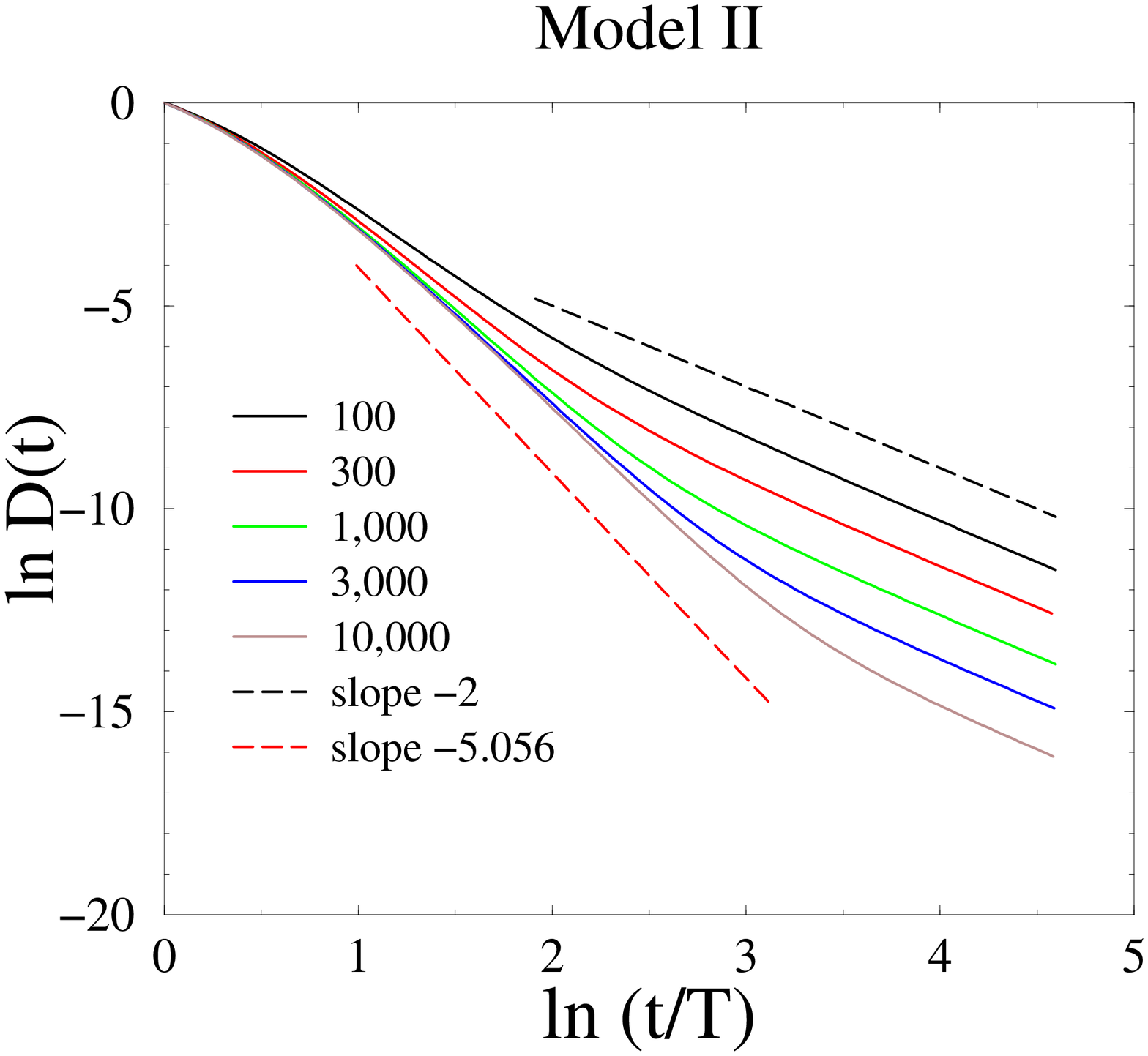}
\caption{\label{aging}
Log-log plot of the total polarisation $D(t)$
against the time ratio $t/T$,
for several durations $T$ of the DC signal (see legends).
The absolute slope of the black dashed lines is $\theta=2$,
while that of the red one for Model~II is $\Theta\approx5.056$.}
\end{center}
\end{figure}

Figure~\ref{aging} shows a log-log plot of $D(t)$ against the time ratio $t/T$,
for both models
in the regime where the duration $T$ of the DC signal
and the observation time $t$ are {\it both} large and comparable.
In the case of Model~I, the data for the longer times exhibit a clear collapse,
indicating a scaling behaviour of the form
\beq
D(t)\approx F(t/T),
\label{faging}
\eeq
which is a signature of (simple) aging~\cite{raging}.
The corresponding scaling function falls off as $F(x)\sim x^{-\theta}$.
For Model~II, the
decay of the total polarisation is more subtle, and
exhibits two successive regimes:
(i) a transient regime,
where $D(t)$ exhibits simple aging in terms of $t/T$,
and falls off rather rapidly;
(ii) an asymptotic regime,
where $D(t)$ falls off with the universal exponent $\theta$,
but does {\it not} obey simple aging.

This qualitative difference between both models is investigated in detail in Appendix~B,
but we give a simple flavour here:
suppose the synapse is in a polarised state
where only the uppermost level is occupied, when the process of forgetting begins.
For Model~I, the polarisation always falls off
with the universal forgetting exponent~$\theta$,
whereas for Model~II it falls off more rapidly,
with a larger transient forgetting exponent~$\Theta$
which depends continuously on $\b$ (see~(\ref{Theta})).
For $\b=0.2$ and $\g=0.5$ we have $\Theta\approx5.056516$.
We are thus led to the following scenario for Model~II:
(i) the bulk of the polarisation profile
is sharply localised around the typical depth~(\ref{nT}) (see Figure~\ref{stef}),
and therefore falls off with the transient exponent $\Theta$;
(ii) the tail of the polarisation profile,
which still has the universal exponential form~(\ref{dinit}) (as long as $T$ is finite),
is responsible for the subsequent universal asymptotic decay.

\section{Fluctuations in default state and signal-to-noise ratio}
\label{stn}

\begin{figure}[!ht]
\begin{center}
\includegraphics[angle=-90,width=.45\linewidth]{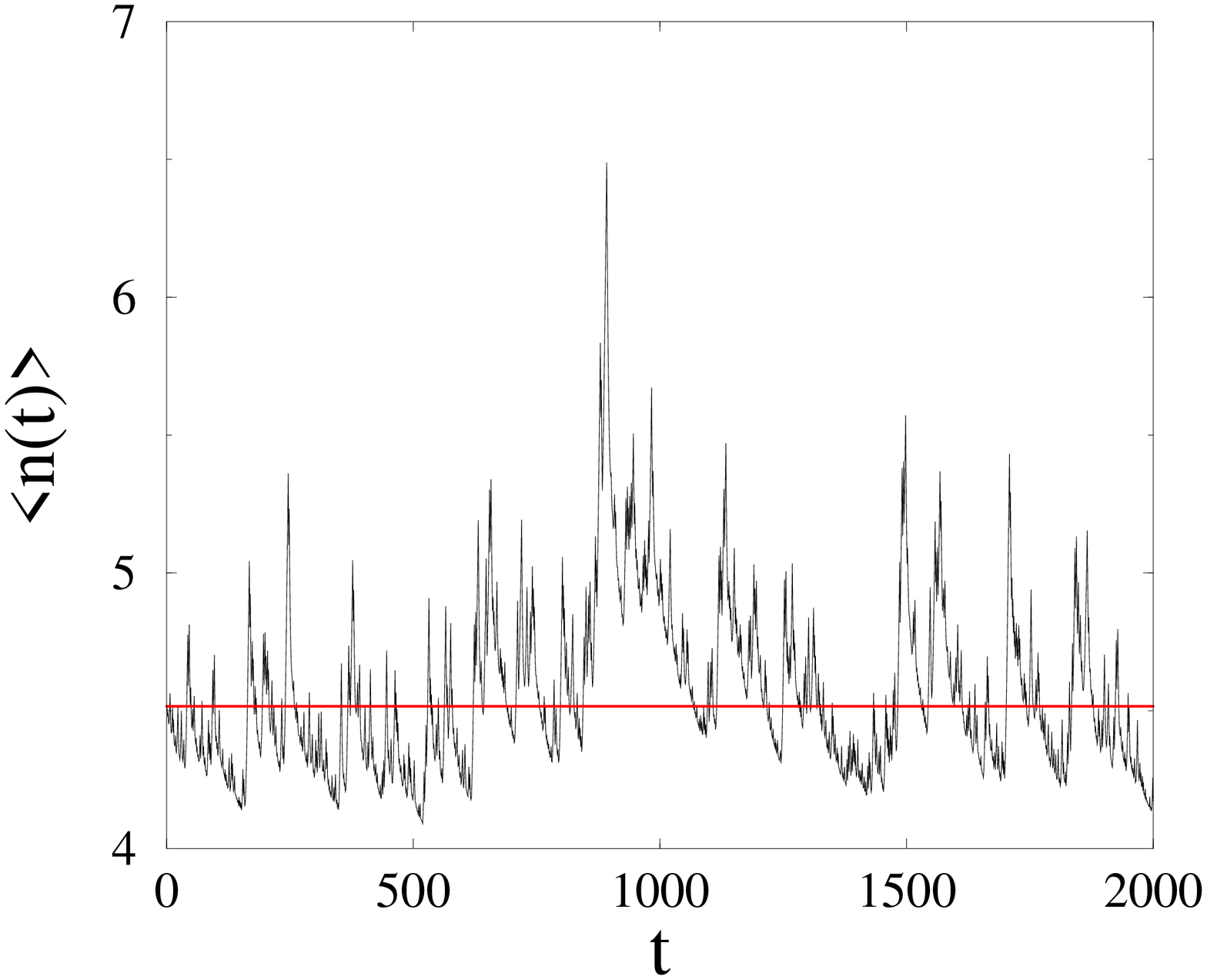}
{\hskip 10pt}
\includegraphics[angle=-90,width=.45\linewidth]{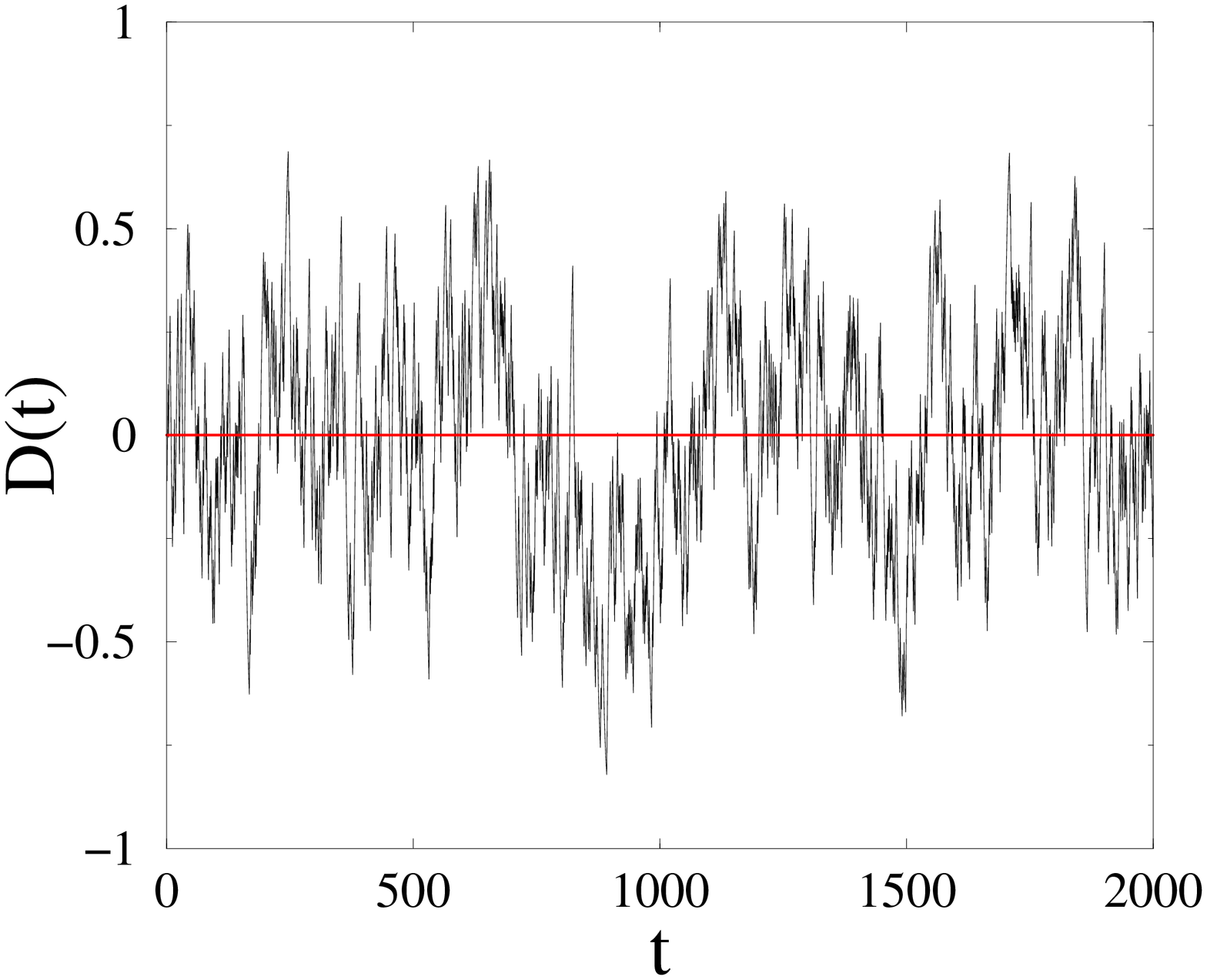}
\caption{\label{ran}
Response of the synapse to a single instance of white-noise random input signal
(Model~I, $\b=0.2$).
Left: mean level depth $\mean{n(t)}$.
Right: total polarisation~$D(t)$.
Red lines: average quantities, characteristic of the default state.}
\end{center}
\end{figure}

The average response of the synapse to a white-noise random input signal
defines its default state, investigated in Section~\ref{default}.
However, there are appreciable dynamical fluctuations around this average,
which are seen on plots (see Figure~\ref{ran})
of the mean level depth $\mean{n(t)}$ (left)
and of the total polarisation $D(t)$ (right)
for Model~I.\footnote{Similar qualitative behaviour
would be obtained for Model~II.}
The average quantities in each case are
shown as red horizontal lines, for ease of comparison.
The mean polarisation vanishes,
whereas the mean level depth is $\mean{n}^\st\approx4.516655$ (see~(\ref{meanstat})).

The large fluctuations observed in both quantities
are due to the occurrence of long ordered subsequences (patches)
of LTP or of LTD events in the input signal
($\eps(t_0+1)=\cdots=\eps(t_0+T)$).
Patches of duration $T$ occur
with exponentially small probabilities $2^{-(T-1)}$, so that
for a total observation time $t$, the largest ordered patch
has $T\approx(\ln 2t)/(\ln 2)$.
For a time of observation
$t=2000$, for example, we can have patches of temporal length as large as
$T\approx12$.
The main effect of, say a long patch of LTP/LTD events, is that
the synapse gets more and more positively/negatively polarised,
with the signal penetrating to ever deeper levels along the appropriate branch.
Such large fluctuations in $D(t)$
are therefore distributed symmetrically around zero,
whereas those in $\mean{n(t)}$ are toward deeper levels.
Clearly, the fluctuations in both quantities should be strongly correlated and
the plots show that they are.

We define the (amplitude) signal-to-noise ratio~$R$ of our models
as the ratio of the instantaneous single LTP signal response $D(1)=\lambda_1\b$
(see~(\ref{done}))
to the standard deviation $D_\rms=\mean{D^2}^{1/2}$
of the spontaneous fluctuations around the default state:
\beq
R=\frac{D(1)}{D_\rms}=\frac{\lambda_1\b}{\mean{D^2}^{1/2}}.
\eeq

\begin{figure}[!ht]
\begin{center}
\includegraphics[angle=-90,width=.45\linewidth]{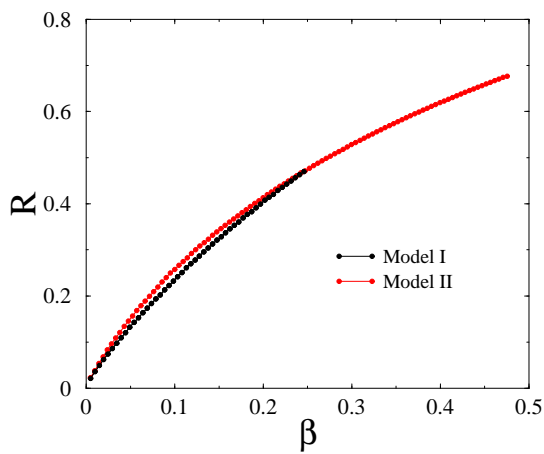}
\caption{\label{r}
Plot of the signal-to-noise ratio $R$ of both models
against $\b\le\b_\max(\g)$ (see~(\ref{bI}),~(\ref{bII})).}
\end{center}
\end{figure}

Figure~\ref{r} shows a plot of the signal-to-noise ratio thus defined,
against $\b$, for both models.
The mean squared polarisation $\mean{D^2}$
is measured by numerically evaluating the response of our models
to a very long sequence of white noise.
Both datasets essentially exhibit the same monotonic dependence on $\b$.
They seem to obey the scaling behaviour $R\sim\sqrt{\b}$ at small $\b$,
suggested by the forthcoming analysis
of the limiting regime $\xis\to0$ (see~(\ref{rzero})).
Conversely, $R$ is maximal at $\b_\max(\g)$,
and this maximal value is essentially determined
by the range $\b_\max(\g)$ of allowed values of $\b$ (see~(\ref{bI}),~(\ref{bII})).

For Model~I with $\xis=\xid=5$, the signal-to-noise ratio $R$ reaches
its global maximum over $\b$ and $\g$,
i.e., $R_\max\approx0.645$, at $\g=1$, i.e., at point G (see Figure~\ref{domains}).
For Model~II, the global maximum
$R_\max\approx0.951$ is reached in the $\g\to0$ limit,
i.e., again at point G.

Optimising the signal-to-noise ratio even further
necessitates allowing the lengths~$\xis$ and $\xid$ to vary;
$R$ is observed to reach its absolute maximum $R=1$ in the $\xis\to0$ limit.
In this regime, the polarisation reads $D(t)=D_0(t)=Q_0(t)-P_0(t)$,
and it is governed by the following simple dynamical equation
\beq
D(t+1)=(1-\b)D(t)+\b\eps(t+1)
\eeq
for both models.
In the stationary state for a white-noise input, we thus~have
\beq
\mean{D^2}=\b^2\sum_{k\ge0}(1-\b)^{2k}=\frac{\b}{2-\b},
\eeq
and finally
\beq
R=\sqrt{\b(2-\b)}.
\label{rzero}
\eeq
The signal-to-noise ratio thus obeys a quarter-of-a-circle law
as a function of $\b$, and attains its absolute maximum $R=1$ at $\b=1$,
irrespective of anything else.

This extreme $\xis\to0$ regime is however of little interest,
as all the action takes place in the uppermost level ($n=0$),
so that metaplasticity is lost.

\section{Response to a variety of input signals}
\label{variety}

In order to examine the storage of memories in the general case, we now examine
the response of our two synapse models to a variety of types
of time-dependent input signals.
As already mentioned in the Introduction,
this section completes the systematic study of our models
viewed from a physicist's perspective as signal processing units.

\subsection{AC signal}
\label{secac}

An AC signal is a perfect alternation of LTP and LTD events,
represented by the input
\beq
\eps(t)=(-1)^t.
\label{acdef}
\eeq

After a short transient, the synapse reaches a stationary state,
where the occupation probabilities keep oscillating in phase
with the input signal, according to
\beq
\matrix{
t\ \hbox{even}\hfill&(\eps(t)=+1):\hfill&P_n(t)=A_n,\hfill&Q_n(t)=B_n,\hfill\cr
t\ \hbox{odd}\hfill&(\eps(t)=-1):\quad\hfill&P_n(t)=B_n,\hfill&Q_n(t)=A_n.\hfill
}
\eeq
The staggered probabilities $A_n$ and $B_n$ are given by the normalised solution
of the following equations:

\cas
Model~I:
\nopagebreak

\beq
\matrix{
A_n=(1-\a_n-\b_n)B_n+\a_{n+1}B_{n+1},\hfill\cr
B_n=(1-\g_n)A_n+\g_{n-1}A_{n-1}+\delta_{n0}\w B,\hfill
}
\label{acI}
\eeq
with
\beq
\w B=\sum_{n\ge0}\b_nB_n.
\eeq

\cas
Model~II:
\nopagebreak

\beq
\matrix{
A_n=(1-\a_n-\b_n)B_n+\a_{n+1}B_{n+1},\hfill\cr
(1-\b_n)B_n=(1-\g_n)A_n+\g_{n-1}A_{n-1}.\hfill
}
\label{acII}
\eeq
The staggered polarisation of the stationary state reads
\beq
D^*=\lim_{t\to\infty}\left(\eps(t)D(t)\right)=\sum_{n\ge0}(B_n-A_n).
\eeq
This quantity starts increasing linearly with $\b$, as
\beq
D^*\approx\lambda_\ac\b,
\label{daclin}
\eeq
irrespective of the model, provided parameters are the same.
In the $\b\to0$ limit,~(\ref{acI}) and~(\ref{acII})
indeed simplify to the same equations
\beq
\matrix{
A_n^\zero=(1-\a_n)B_n^\zero+\a_{n+1}B_{n+1}^\zero,\hfill\cr
B_n^\zero=(1-\g_n)A_n^\zero+\g_{n-1}A_{n-1}^\zero,\hfill
}
\label{aczero}
\eeq
whose normalised solution $A_n^\zero$, $B_n^\zero$ is therefore
model-independent.
We have then
\beq
\lambda_\ac=\sum_{n\ge0}\e^{-n\mud}B_n^\zero.
\eeq
For the parameters~(\ref{pars}) this gives $\lambda_\ac\approx0.329712$.

\begin{figure}[!ht]
\begin{center}
\includegraphics[angle=-90,width=.45\linewidth]{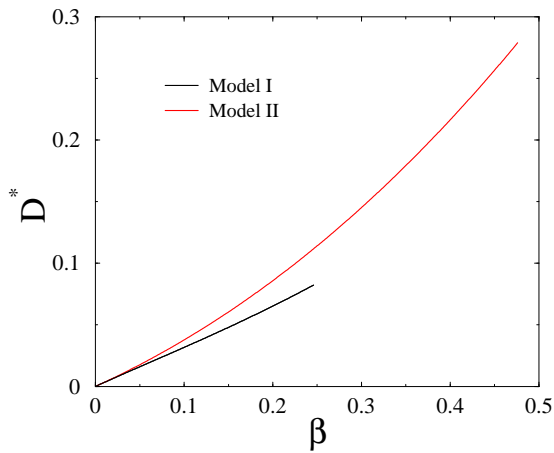}
\caption{\label{ac}
Plot of the stationary staggered polarisation $D^*$ of both models
submitted to an AC signal, against $\b\le\b_\max(\g)$ (see~(\ref{bI}),~(\ref{bII})).}
\end{center}
\end{figure}

Figure~\ref{ac} shows a plot of $D^*$ against $\b$ for both models.
For each model,~$\b$ is limited by $\b_\max(\g)$.
Once again we see that the staggered polarisation reaches larger values for Model~II,
mainly as a consequence of the larger range of allowed values of $\b$.

\subsection{Coloured random signal}
\label{seccolour}

We next consider a coloured random input signal
defined by the following rule:
\beq
\eps(t+1)=\left\{\matrix{
{\hskip 9.25pt}\eps(t)\hfill&\hbox{with probability }r,\hfill\cr
-\eps(t)\quad\hfill&\hbox{with probability }1-r,\hfill
}\right.
\label{coldef}
\eeq
with $\eps(1)=+1$ for definiteness.

The persistence probability $r$ allows this coloured random signal
to interpolate between several situations described above:

\cas
the DC signal investigated in Section~\ref{secdc} is recovered for $r=1$,

\cas
the AC signal investigated in Section~\ref{secac} is recovered for $r=0$,

\cas
the white-noise signal investigated in Sections~\ref{default} and~\ref{stn}
is recovered for $r=\half$.

The correlation function of the signal~(\ref{coldef}) is
\beq
S(t)=\mean{\eps(t_0)\eps(t_0+t)}=(2r-1)^t\qquad(t\ge0).
\eeq
The coloured signal is therefore positively correlated, or persistent, for
$\half<r<1$.
The corresponding characteristic time
\beq
\tau=\frac{1}{\abs{\ln(2r-1)}}
\eeq
diverges near the DC limit ($r\to1$) as $\tau\approx1/(2(1-r))$.
The signal is anti-persistent, with oscillating correlations, for $0<r<\half$.

\begin{figure}[!ht]
\begin{center}
\includegraphics[angle=-90,width=.45\linewidth]{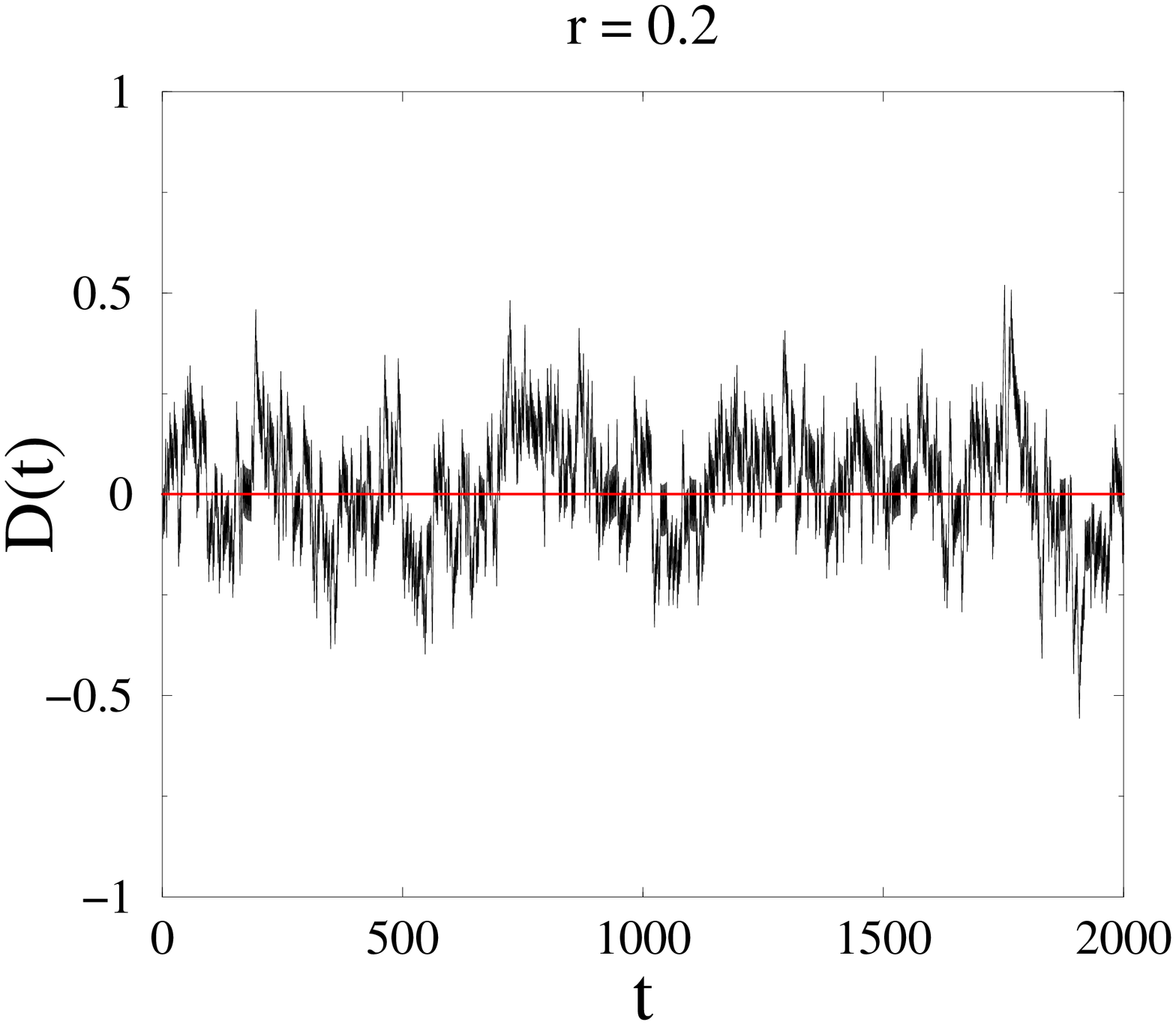}
{\hskip 10pt}
\includegraphics[angle=-90,width=.45\linewidth]{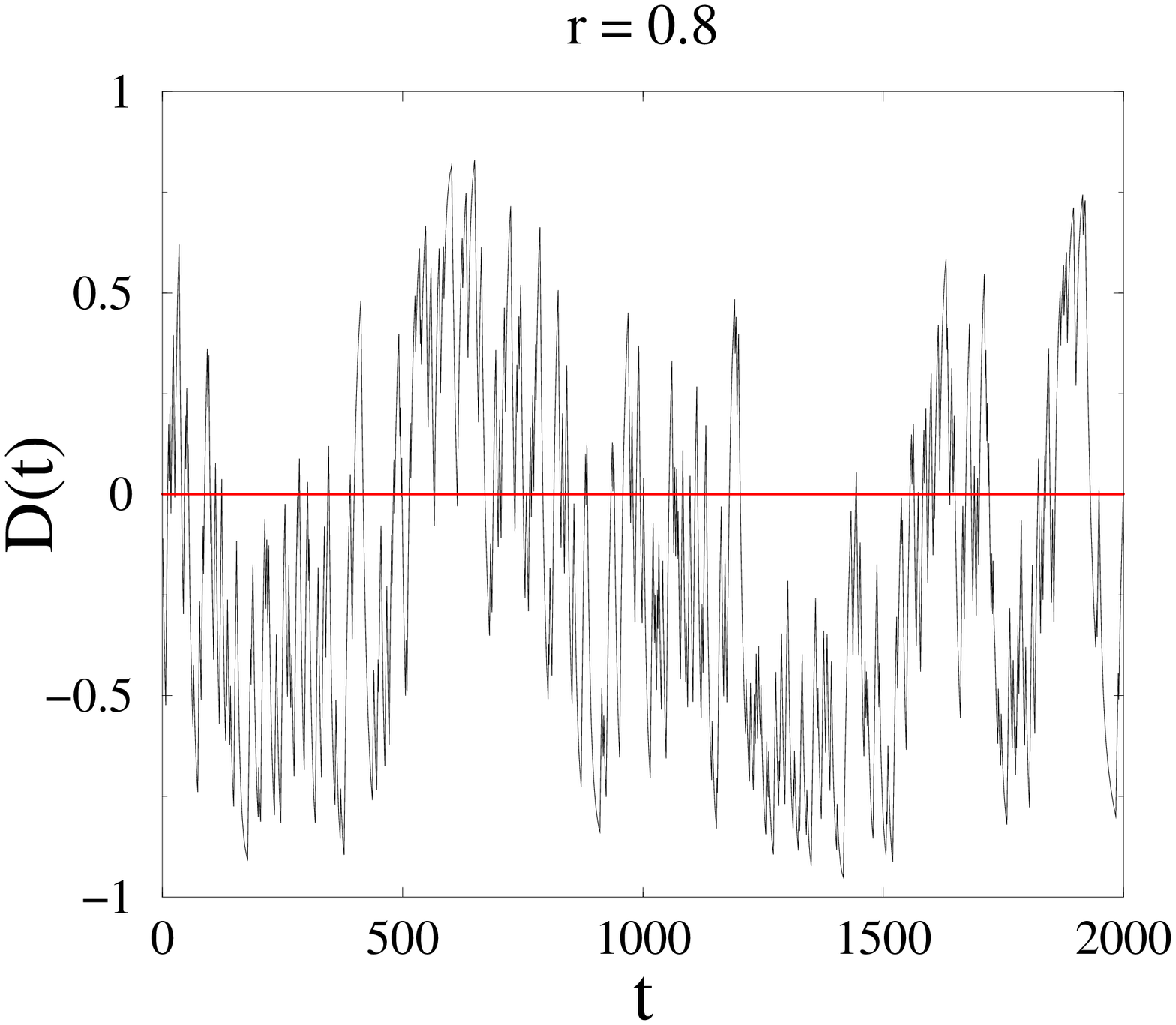}
\caption{\label{col28}
Plot of the total polarisation $D(t)$ in response
to a single realisation of coloured random input (Model~I, $\b=0.2$).
Left: an anti-persistent case ($r=0.2$).
Right: a persistent case ($r=0.8$).}
\end{center}
\end{figure}

The synapse submitted to a coloured random input signal
reaches a fluctuating stationary state after a relatively short transient.
For a given realisation, it exhibits strong dynamical fluctuations
which are qualitatively similar to those shown in Figure~\ref{ran}.
Figure~\ref{col28} shows plots of the total polarisation $D(t)$
for two typical realisations of coloured random signal,
in an anti-persistent case ($r=0.2$, left)
and in a persistent case ($r=0.8$, right).
Both the amplitude and the correlation time of the fluctuations
are observed to increase with~$r$, as might be expected.

\begin{figure}[!ht]
\begin{center}
\includegraphics[angle=-90,width=.45\linewidth]{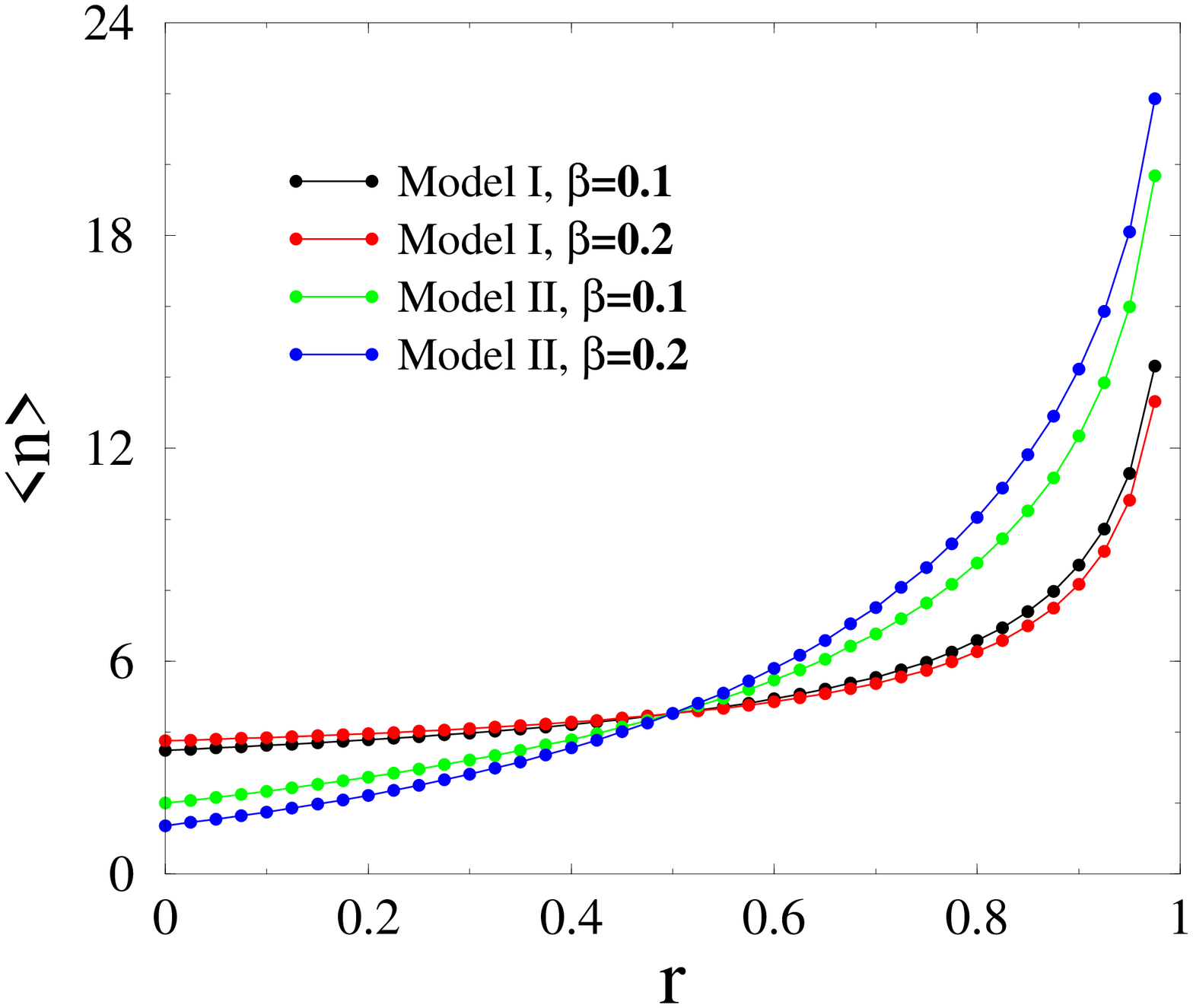}
{\hskip 10pt}
\includegraphics[angle=-90,width=.45\linewidth]{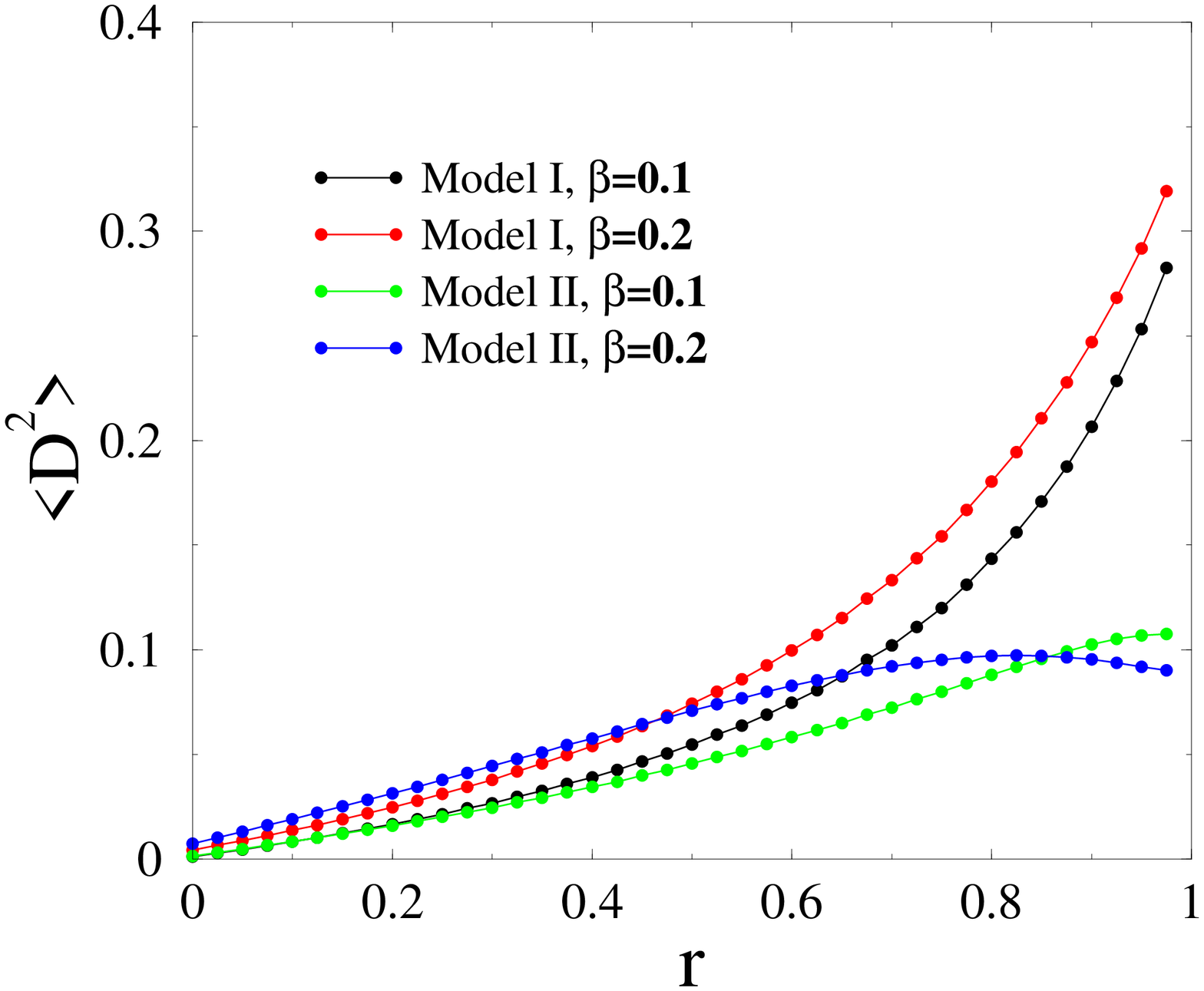}
\caption{\label{colour}
Plot of the stationary values
of the mean depth $\mean{n}$ (left)
and of the mean squared polarisation $\mean{D^2}$ (right),
against the persistence probability~$r$,
for both models with $\b=0.1$ and 0.2.}
\end{center}
\end{figure}

Figure~\ref{colour} shows a plot of (numerically measured) stationary values
of the mean depth $\mean{n}$ (left)
and of the mean squared polarisation $\mean{D^2}$ (right),
for both models with $\b=0.1$ and 0.2
and a varying persistence probability~$r$.
The mean depth starts from its lowest value in the $r\to0$ limit,
i.e., for the AC signal.
It increases smoothly as a function of $r$, and diverges logarithmically as
$\mean{n}\approx\xid\ln\tau\approx\xid\abs{\ln(1-r)}$
near the DC limit ($r\to1$).\footnote{This logarithmic law can be derived in
the same spirit as~(\ref{log}) and~(\ref{nT}).}
All the curves cross at the white-noise point ($r=\half$),
where the result~(\ref{meanstat}) holds irrespective of the model and of its parameters.
The dependence of the mean depth on the persistence probability~$r$
is far more pronounced for Model~II than for Model~I.
The behaviour of the mean squared polarisation $\mean{D^2}$
provides another appreciable difference between the two models.
In both cases it starts increasing as a function of $r$,
from a very small value in the $r\to0$ limit.
Its behaviour as $r\to1$ is however very different in both models.
The mean squared polarisation keeps steadily increasing in the case of Model~I,
whereas its increase is much less pronounced for Model~II, even becoming
non-monotonic at high enough $\b$.

This qualitative difference between the responses of both models
to highly persistent random signals can be related
to the difference in their transient responses,
investigated in Section~\ref{tplf}.
In Model~II,
the low-frequency components of the memory lie slightly deeper within the synapse.
More importantly, they relax much faster than in Model~I,
as their falloff can be characterised by
a larger, non-universal exponent $\Theta$.

\subsection{Oscillatory signal}

The last case we consider is that of an oscillatory input signal, which
consists of alternating long blocks of LTP and LTD signals
of length $T$ time steps, i.e.,
\beq
\eps(t)=(-1)^{\Int(t/T)},
\label{oscdef}
\eeq
where $\Int$ denotes the integer part.

\begin{figure}[!ht]
\begin{center}
\includegraphics[angle=-90,width=.45\linewidth]{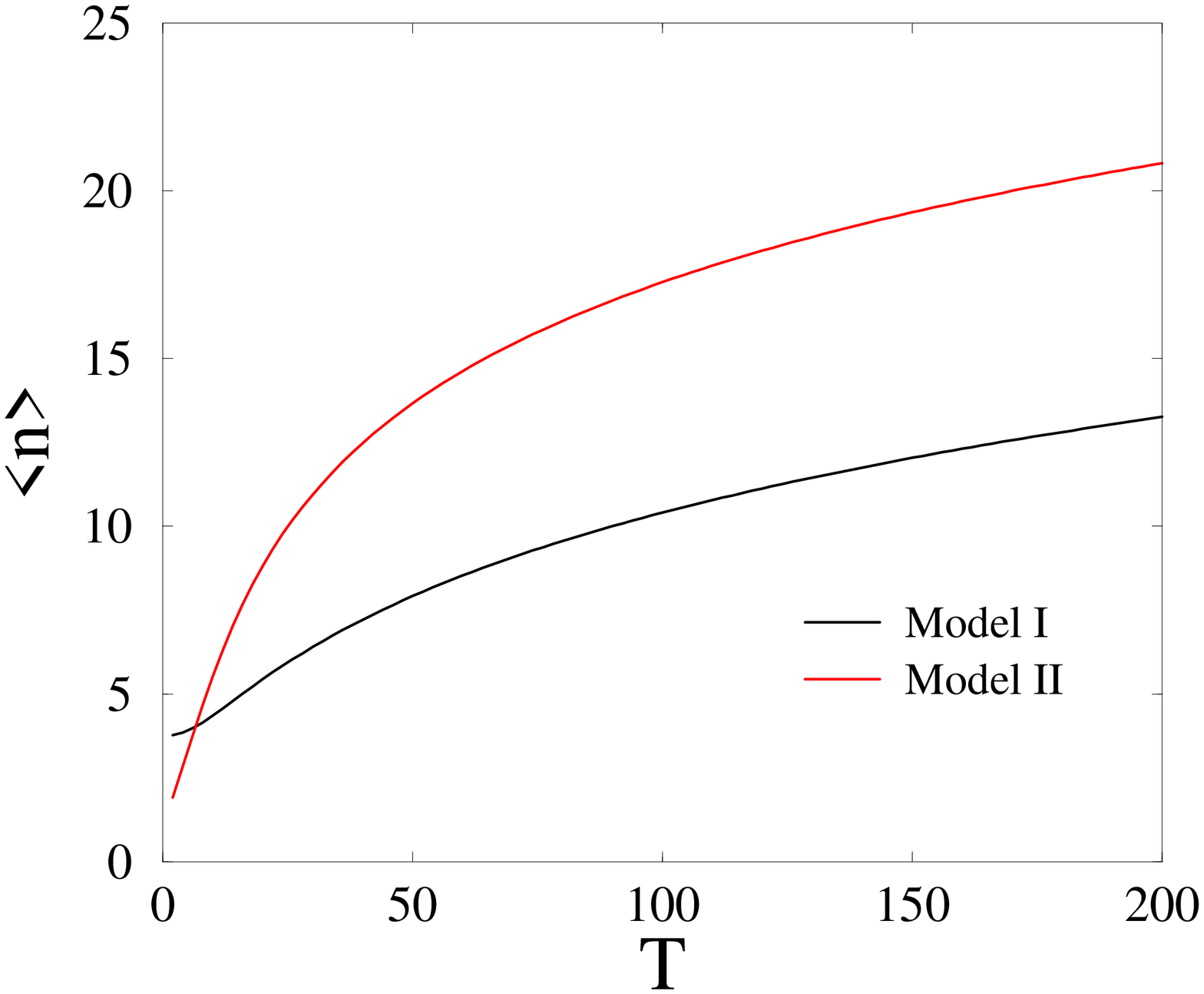}
{\hskip 10pt}
\includegraphics[angle=-90,width=.45\linewidth]{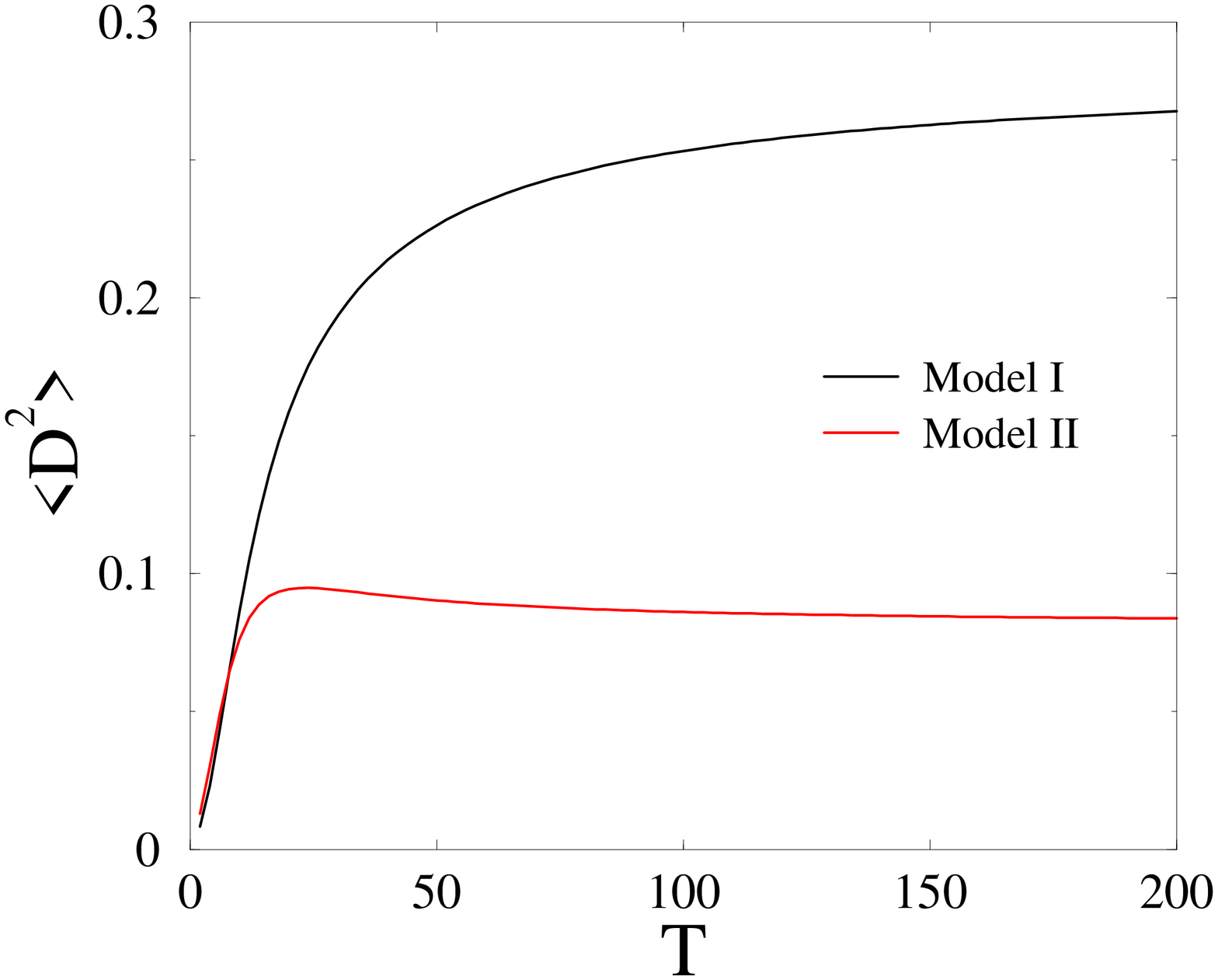}
\caption{\label{per}
Plot of the stationary values
of the mean depth $\mean{n}$ (left)
and of the mean squared polarisation $\mean{D^2}$ (right)
for both models with $\b=0.2$,
against the half-period~$T$ of the oscillatory signal.}
\end{center}
\end{figure}

After a relatively short transient regime,
the synapse converges toward a periodic state,
where the polarisation and other quantities
oscillate with the period $2T$ of the input signal.
Figure~\ref{per} shows a plot of the values
of the mean depth $\mean{n}$ (left)
and of the mean squared polarisation $\mean{D^2}$ (right),
averaged over one period in the stationary state of the synapse,
for both models with $\b=0.2$, against the half-period $T$ of the oscillatory signal.

These data corroborate the observations made in Section~\ref{seccolour}.
The dependence of the mean depth on $T$
is again steeper for Model~II than for Model~I.
The data for both models are however compatible with the common logarithmic
asymptotic growth law $\mean{n}\approx\xid\ln T$.
The mean squared polarisation $\mean{D^2}$ is observed
to increase monotonically as a function of $T$ for Model~I,
whereas for Model~II it reaches a maximum and then smoothly decreases.

\section{Discussion}
\label{discussion}

In this paper, we have provided the first thorough analysis of
a single synapse including the effects of metaplasticity.
We used two models:
Model~I is an extension of the original cascade model proposed by Fusi~\etal~\cite{fusi},
whereas Model~II, of our own invention, has a different architecture.

Our intention was, apart from the thorough quantification of earlier
ideas~\cite{amfuss}, the isolation of the mechanisms responsible for the
storage of memories, and the differentiation of short-
and long-term memories in response to a range of signal types.
In the structure of the models we analysed, long-term memories
were stored at greater `depths', and therefore relatively immunised to the
constant bombardment of white noise in the upper levels,
which forms our everyday experience.
The difference between the two models lies
chiefly in the mechanism of response of the synapse to a flip in sign of the input signal.
In Model~I, such flips tend to cause the memory
trace to be dislodged to become a short-term memory of the opposite kind,
while in Model~II, changes in long-term memories are allowed to
be more persistent, remaining at low level depths.

Most remarkably, the same asymptotic power-law forgetting
(with universal exponent $\theta$)
is manifested in both models, independent of their architecture.
However, the aging behaviour of the models is rather different:
Model~I manifests simple aging,
whereas Model~II may have a long transient regime,
where the polarisation falls off more rapidly
(with a larger, non-universal and $\b$-dependent transient forgetting exponent $\Theta$),
before the asymptotic power-law forgetting takes over.

The behaviour of both models has been further illustrated
by their responses to a range of input signals.
The two observables we focused on were the mean depth of a particular memory
trace, and its polarisation.
Our observations suggest that Model~II allows in general
for a slightly greater penetration of signals.
The long-term memories thus created are however rather weaker than in the case of Model~I.
This weakening is to be put in perspective
with the existence of the non-universal transient exponent,
which is studied at length in Appendix~B.
Qualitatively speaking, it appears
that the changing of `opinions' represented by the two poles of a synapse at a
relatively deep level (which is possible in Model~II) has the effect of weakening the
strength of a memory trace, far more than when contradictions are resolved by
simply disposing of them in the short-term memories of the opposite pole.

Our results provide the first prediction of the exponent
of power-law forgetting at the level of a single synapse:
the intensive analysis of these relatively simple models could help to start
theoretical work on more complex architectures,
since of course real-life forgetting relies not just on individual synapses,
but on their connections to each other and to neurons.
Possible extensions of this work might involve the coupling~\cite{gmam}
of multiple synapses of the type presented above,
or include the effect of correlated signals~\cite{rt};
increasing correlations in these ways might enhance the competition
between the bulk and the tails of the signals deep within a metaplastic synapse.

\ack

We warmly thank Nicolas Brunel for very fruitful discussions
and Mark van Rossum for interesting correspondence.
AM thanks the Institut de Physique Th\'eorique, where much of this work was carried out,
for its customary gracious hospitality during her visits.

\appendix

\section{The logarithmic walker}

The problem of the logarithmic walker is defined as follows.
A particle lives on the semi-infinite chain,
whose sites are labelled by the positive integers ($n=0,1,\dots$).
At each time step, if sitting at site $n$, the particle may hop to the right
($n\to n+1$) with exponentially decaying probabilities $\e^{-n\mu}$.
This model can be alternatively thought of as describing
a discrete-time pure birth process,
which has already been considered in~\cite{grimmett}.

If we assume for definiteness that the particle starts from the origin at time $t=0$,
the probability $p_n(t)$ for the particle to be at site $n$ at time $t$
obeys the recursion
\beq
p_n(t+1)=(1-\e^{-n\mu})p_n(t)+\e^{-(n-1)\mu}p_{n-1}(t),
\label{plw}
\eeq
with initial condition $p_n(0)=\delta_{n0}$.

\begin{figure}[!ht]
\begin{center}
\includegraphics[angle=-90,width=.45\linewidth]{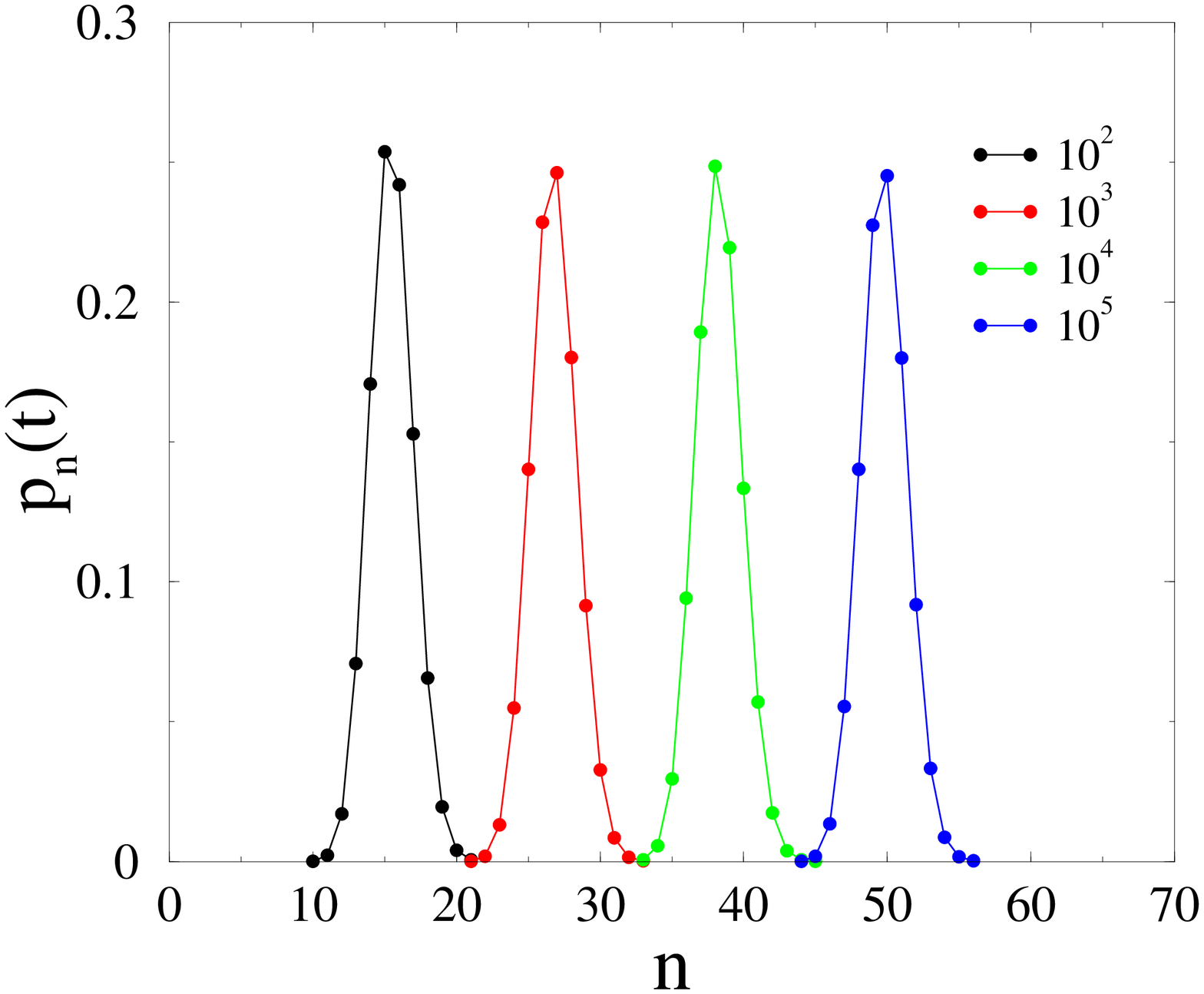}
\caption{\label{front}
Plot of the probability profile $p_n(t)$ of the logarithmic walker
against position $n$, for $\mu=0.2$ and several times $t$ (see legend).}
\end{center}
\end{figure}

Figure~\ref{front} shows a plot of the probability profile $p_n(t)$ against $n$,
for $\mu=0.2$ and several times~$t$.
The profile is observed to form a peak around a well-defined mean
position $\mean{n(t)}$.
As time goes on, the profile keeps its shape,
while the mean position exhibits a very slow growth.

A first heuristic approach to estimate this growth law
consists in writing down the dimensional estimate $\e^{-n\mu}\,t\sim1$,
hence the result
\beq
\mean{n(t)}\approx\frac{\ln t}{\mu},
\label{log}
\eeq
and the name, {\it logarithmic walker}.

A more precise approach consists of writing down
the following dynamical equation
for the mean position of the walker at time $t$:
\beq
\mean{n(t+1)}-\mean{n(t)}=\sum_{n\ge0}\e^{-n\mu}p_n(t).
\label{dmean}
\eeq
For a localised probability profile,
we thus obtain approximately $\d\mean{n}/\d t\approx\e^{-\mean{n}\mu}$, which yields
\beq
\mean{n(t)}\approx\frac{\ln(1+\mu t)}{\mu},
\label{log2}
\eeq
in agreement with~(\ref{log}).

Turning to more quantitative analysis,
we look for an asymptotic solution to~(\ref{plw})
in the form of a traveling wave moving on a logarithmic time scale, i.e.,
\beq
p_n(t)\approx F(x),\qquad x=n-\lambda,\qquad\lambda=\frac{\ln t}{\mu}.
\label{appq}
\eeq

It is worth emphasising the difference between the present situation,
where time~$t$ enters the argument $x$ through its logarithm
and with an explicitly known prefactor $1/\mu$,
and the usual situation of ballistic traveling waves,
like e.g.~in the FKPP equation~\cite{fkpp}.
For such traveling waves,
time $t$ is multiplied by an unknown velocity $v$,
whose determination is non-trivial,
and known to be very sensitive to discretization
and other fluctuation effects~\cite{flucs}.

The {\it hull function} $F(x)$ of the traveling wave~(\ref{appq})
is found to obey the linear differential-difference equation
\beq
F'(x)=\mu\e^{-\mu x}(F(x)-\e^\mu F(x-1)).
\label{appf}
\eeq
As a consequence, its Laplace transform
\beq
L_F(s)=\int_{-\infty}^{+\infty}\e^{-sx}\,F(x)\,\d x
\label{lapdef}
\eeq
obeys the difference equation
\beq
sL_F(s)=\mu(1-\e^{-s})L_F(s+\mu),
\label{applap}
\eeq
whose normalised solution reads
\beq
L_F(s)=\frac{1-\e^{-s}}{\mu}\,\Gamma\left(\frac{s}{\mu}\right)\,P(s),
\label{appsol}
\eeq
where $P(s)$ is the infinite product
\beq
P(s)=\prod_{k\ge1}\frac{1-\e^{-s-k\mu}}{1-\e^{-k\mu}}.
\eeq
The latter product has zeros on the semi-infinite lattice of points
$s=-k\mu+2\pi\i l$, with~$k$ and $l$ integers such that $k\ge1$,
whereas~$L_F(s)$ shares these zeros for $l\ne0$ only.

For the time being, let us consider the characteristic function of the position $n$,
i.e., the generating function of the probabilities $p_n(t)$:
\beq
E(u,t)=\mean{\e^{un(t)}}=\sum_{n\ge0}p_n(t)\e^{un}.
\eeq
In the long-time regime,
the traveling-wave form~(\ref{appq}) of the probabilities translates to
\beq
E(u,t)\approx\sum_{n=-\infty}^\infty F(n-\lambda)\e^{un}
=\sum_{l=-\infty}^\infty L_F(2\pi\i l-u)\e^{(u-2\pi\i l)\lambda},
\label{appe}
\eeq
where the right-hand side has been obtained by means of the Poisson summation formula.

Setting $u=0$ in~(\ref{appe}), we obtain unity identically, as expected.
The reason is that we have $L_F(0)=1$, whereas $L_F(2\pi\i l)=0$ for $l\ne0$;
thus, the hull function $F(x)$ has the remarkable property
that the `stroboscoped' sum
equals one for all values of the real variable $\lambda$:
\beq
\sum_{n=-\infty}^\infty F(n-\lambda)=1.
\eeq

The asymptotic behaviour of the mean position $\mean{n(t)}$ can be derived
by expanding the result~(\ref{appe}) to first order in $u$.
We thus obtain
\beq
\mean{n(t)}\approx\frac{\ln t+\euler}{\mu}+\frac12-P'(0)
-\frac{1}{\mu}\sum_{l\ne0}\Gamma\left(\frac{2\pi\i l}{\mu}\right)\e^{-2\pi\i
l\lambda},
\eeq
where $\euler$ denotes Euler's constant.
The sum is the Fourier series
of a periodic function of $\lambda$, with unit period.
These oscillations originate in the discrete nature of the sites of the chain.
They manifest themselves e.g.~in the shape of the probability profile near its top.
Oscillations are however extremely small
for global quantities such as $\mean{n(t)}$.
Their amplitude is essentially given by the first Fourier coefficients ($l=\pm1$),
which are proportional to $\e^{-\pi^2/\mu}$.
For $\mu=0.2$, this amplitude is of order $\e^{-5\pi^2}\sim10^{-22}$,
while for $\mu=1$ it is of order $\e^{-\pi^2}\sim10^{-5}$.

Neglecting these (tiny) oscillations, we are left with
\beq
\mean{n(t)}\approx\frac{\ln t+\euler}{\mu}+\frac12-P'(0).
\label{appn}
\eeq
This expression confirms the estimates~(\ref{log}) and~(\ref{log2}),
and gives an explicit expression for the finite part of the logarithm, where
\beq
P'(0)=\sum_{k\ge1}\frac{1}{\e^{k\mu}-1}
=\int\frac{\d z}{2\pi\i\mu^z}\,\Gamma(z)\zeta(z)^2.
\eeq
The Mellin-Barnes integral representation of the right-hand side,
where $\zeta(z)$ is Riemann's zeta function,
is suited to the derivation
of the expansion of $P'(0)$ in the regime of small~$\mu$.
We thus obtain the rapidly convergent expansion
\beq
P'(0)=\frac{\euler-\ln\mu}{\mu}+\frac{1}{4}
-\frac{\mu}{144}-\frac{\mu^3}{86\,400}+\cdots,
\eeq
hence
\beq
\mean{n(t)}\approx\frac{\ln\mu t}{\mu}+\frac{1}{4}
+\frac{\mu}{144}+\frac{\mu^3}{86\,400}+\cdots
\eeq
The full leading term was already correctly predicted in~(\ref{log2}).

A similar treatment of the second moment $\mean{n(t)^2}$
demonstrates that the variance of the position saturates to the asymptotic value
\beq
\var n=\lim_{t\to\infty}\left(\mean{n(t)^2}-\mean{n(t)}^2\right)
=\frac{\pi^2}{6\mu^2}+\frac{1}{12}-K,
\label{appvar}
\eeq
again up to negligibly small periodic oscillations, with
\beq
K=P'(0)^2-P''(0)=\sum_{k\ge1}\frac{\e^{k\mu}}{(\e^{k\mu}-1)^2}
=\int\frac{\d z}{2\pi\i\mu^z}\,\Gamma(z)\zeta(z)\zeta(z-1).
\eeq
We thus obtain
\beq
K=\frac{\pi^2}{6\mu^2}-\frac{1}{2\mu}+\frac{1}{24}+\cdots,
\eeq
hence
\beq
\var n=\frac{1}{2\mu}+\frac{1}{24}+\cdots
\eeq
In both above expressions the dots stand for an exponentially small contribution,
proportional to $\e^{-4\pi^2/\mu}$.

To close, let us come back to the form of the hull function $F(x)$,
which describes the asymptotic shape of the probability profile.
The decay of the hull function at both ends is faster than exponential,
since its Laplace transform $L_F(s)$, given in~(\ref{appsol}),
is an entire function, i.e., it is analytic in the whole~$s$ plane.
The decay of $F(x)$ as $x\to\pm\infty$
can be derived by inserting the asymptotic behaviour of $L_F(s)$
as $s\to\mp\infty$ in the inverse Laplace formula
\beq
F(x)=\int\frac{\d s}{2\pi\i}\,\e^{sx}\,L_F(s).
\label{appilt}
\eeq
For $s\to+\infty$, we have $L_F(s)\approx\Gamma(s/\mu)/\mu$.
We thus obtain a double exponential~decay:
\beq
F(x)\approx\exp(\e^{-\mu x})\qquad(x\to-\infty).
\eeq
For $s\to-\infty$, with exponential accuracy, we have
$L_F(s)\sim\e^{-s}L_F(s+\mu)$,
so that $L_F(s)\sim\e^{s^2/(2\mu)}$.
The hull function therefore falls off as a Gaussian:
\beq
F(x)\sim\e^{-\mu x^2/2}\qquad(x\to+\infty).
\eeq
Finally, in the regime of small $\mu$,
where the variance $\var n\approx1/(2\mu)$ is large,
the whole hull function is nearly Gaussian:
\beq
F(x)\approx\sqrt{\frac{\mu}{\pi}}\,
\exp\left(-\mu\left(x-(\ln\mu)/\mu\right)^2\right).
\label{appgau}
\eeq

\section{Transient behaviour}

This Appendix is devoted to the transient behaviour of our models.
Our main goal is to show that the transient responses of both models
are qualitatively different.
The polarisation of Model~II
exhibits a non-universal power-law decay,
with a transient exponent $\Theta$ which depends continuously on $\b$
(see~(\ref{Theta})),
whereas the polarisation of Model~I always falls off
with the universal exponent $\theta$ (see~(\ref{theta})).

To probe this further,
we analyse the transient average response of the synapse to a white-noise random input.
We shift time so that $t=0$ is the beginning of the forgetting period.
For simplicity and without loss of generality,
we caricature transient effects by choosing an initial state
such that the transient regime will last forever.
More specifically, we assume that the synapse is prepared in a totally polarised state
living entirely on the uppermost level: $P_n(0)=0$, $Q_n(0)=\delta_{n0}$.
We will successively consider the level occupation probabilities
and the level-resolved and total polarisations.

\subsection*{Level occupation probabilities}
\label{appocc}

We begin with the level occupation probabilities $S_n(t)$.
The following scenario is expected in the long-time regime:
the $S_n(t)$ should converge rather fast to their stationary values $S_n^\st$
(see~(\ref{sdstat})) at moderate level depths,
whereas their values at very deep levels should fall off more rapidly,
as these are unaffected by the random input.

From a quantitative viewpoint, along the lines of~(\ref{appq}),
we look for an asymptotic long-time solution to~(\ref{meanI}) or~(\ref{meanII})
for the $S_n(t)$
in the form of a traveling wave (front) moving on a logarithmic time scale:
\beq
S_n(t)\approx S_n^\st\,\Phi(x),\qquad x=n-\xid\ln\g t.
\eeq
The scaling function $\Phi(x)$
is expected to decrease from 1 in the $x\to-\infty$ limit
to 0 in the $x\to+\infty$ limit.
Both models have to be dealt with separately.

\cas
Model~I:
\nopagebreak

The function $\Phi(x)$ describing the front obeys the equation
\beqa
-2\g\Phi'(x)&=&\mud\e^{-\mud x}\Big(\a\e^{-\mus}\Phi(x+1)\nonumber\\
&-&(\a\e^{\mud}+\b+\g)\Phi(x)+\g\e^{\mus+\mud}\Phi(x-1)\Big).
\eeqa
Along the lines of Appendix~A,
we introduce the Laplace transform $L_\Phi(s)$ of $\Phi(x)$,
which obeys the functional equation
\beq
-2\g sL_\Phi(s)=\mud
(\a\e^{-\mus+\mud+s}-\a\e^\mud-\b-\g+\g\e^{\mus-s})L_\Phi(s+\mud).
\label{phieq}
\eeq
The expected behaviour of $\Phi(x)$ implies that $L_\Phi(s)$ is analytic for $s<0$
and has a simple pole at $s=0$ with unit residue (i.e.,
$\lim_{s\to0}\left(s L_\Phi(s)\right)=1$).
The property that $L_\Phi(s)$ has no pole at $s=-\mud$
implies that the expression inside the parentheses on the right-hand side
of~(\ref{phieq}) vanishes for $s=-\mud$.
We thus recover~(\ref{abgI}).
The function $L_\Phi(s)$
can be given as an explicit expression similar to~(\ref{appsol}),
involving two infinite products,
which will not be needed in the following.

\cas
Model~II:
\nopagebreak

The analysis is very similar.
The function $\Phi(x)$ obeys the equation
\beqa
-2\g\Phi'(x)&=&\mud\e^{-\mud x}\Big(\a\e^{-\mus}\Phi(x+1)\nonumber\\
&-&(\a\e^{\mud}+\g)\Phi(x)+\g\e^{\mus+\mud}\Phi(x-1)\Big).
\eeqa
The Laplace transform $L_\Phi(s)$ obeys
\beq
-2\g sL_\Phi(s)=\mud(1-\e^{\mus-s})(\a\e^{-\mus+\mud+s}-\g)L_\Phi(s+\mud).
\eeq
The absence of a pole at $s=-\mud$ implies $\g=\a\e^{-\mus}$.
We thus recover~(\ref{abgII}).

\subsection*{Level-resolved and total polarisations}
\label{apppol}

We now turn to the analysis of the level-resolved polarisations $D_n(t)$
and of the total polarisation $D(t)$.
We anticipate a power-law decay in the long-time regime.
Thus, we look for an asymptotic solution to~(\ref{meanI}) or~(\ref{meanII})
for the~$D_n(t)$ in the form of a power-law decay,
with a positive exponent $\Theta$,
which multiplies a logarithmic front:
\beq
D_n(t)\sim t^{-\Theta}\,\Psi(x),\qquad x=n-\xid\ln\g t.
\eeq
The scaling function $\Psi(x)$
is expected to decrease fast enough as $x\to+\infty$,
in such a way that the total polarisation of the synapse
also falls off as $D(t)\sim t^{-\Theta}$.
Both models again have to be dealt with separately.

\cas
Model~I:
\nopagebreak

The function $\Psi(x)$ obeys the equation
\beqa
-2\g(\Psi'(x)+\Theta\mud\Psi(x))&=&\mud\e^{-\mud x}\Big(\a\Psi(x+1)\nonumber\\
&-&(\a\e^{\mud}+\b+\g)\Psi(x)+\g\e^\mud\Psi(x-1)\Big).
\eeqa
The Laplace transform $L_\Psi(s)$ of $\Psi(x)$
obeys the functional equation
\beqa
-2\g(s+\Theta\mud)L_\Psi(s)&=&\mud\Big(\a\e^{\mud+s}\nonumber\\
&-&\a\e^\mud-\b-\g+\g\e^{-s}\Big)L_\Psi(s+\mud).
\eeqa
The fast decay of $\Psi(x)$ as $x\to+\infty$
implies that $L_\Psi(s)$ is analytic at least for $s<0$.
The absence of a pole at $s=-\Theta\mud$ yields
$\a\e^{(1-\Theta)\mud}-\a\e^\mud-\b-\g+\g\e^{\Theta\mud}=0$.
Using~(\ref{abgI}), this condition simplifies to
\beq
(\e^{\Theta\mud}-\e^{\mus+\mud})(\g\e^{\Theta\mud}-\a\e^{-\mus})=0.
\label{equals0}
\eeq
The vanishing of the first factor\footnote{The second factor
of~(\ref{equals0}) is positive, as a consequence of~(\ref{abgI}).}
leads to the simple result that $\Theta$
is identical to the universal forgetting exponent $\theta$ (see~(\ref{theta})).
We have thus shown that Model~I exhibits a remarkably universal power-law forgetting.

\cas
Model~II:
\nopagebreak

\begin{figure}[!ht]
\begin{center}
\includegraphics[angle=-90,width=.45\linewidth]{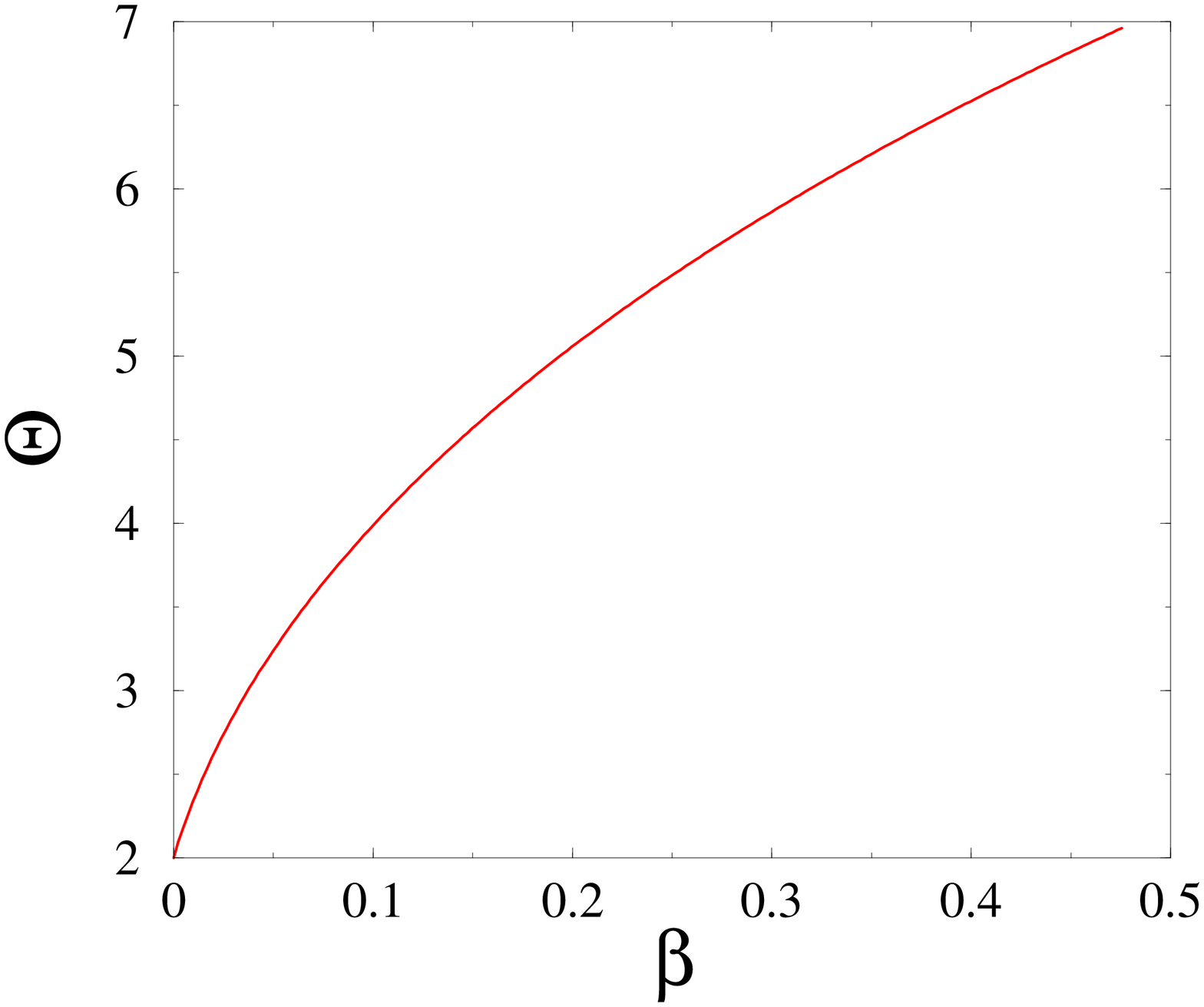}
\caption{\label{Th}
Plot of the non-universal transient forgetting exponent $\Theta$
of Model~II with the parameters~(\ref{pars}),
against $\b\le\b_\max(\g)$.}
\end{center}
\end{figure}

The analysis is similar, although it leads to a very different outcome.
The function $\Psi(x)$ obeys the equation
\beqa
-2\g(\Psi'(x)+\Theta\mud\Psi(x))&=&\mud\e^{-\mud x}\Big(\a\Psi(x+1)\nonumber\\
&-&(\a\e^{\mud}+2\b+\g)\Psi(x)+\g\e^\mud\Psi(x-1)\Big).
\eeqa
The Laplace transform $L_\Psi(s)$ of $\Psi(x)$
obeys the functional equation
\beqa
-2\g(s+\Theta\mud)L_\Psi(s)&=&\mud\Big(\a\e^{\mud+s}\nonumber\\
&-&\a\e^\mud-2\b-\g+\g\e^{-s}\Big)L_\Psi(s+\mud).
\eeqa
The absence of a pole at $s=-\Theta\mud$ yields
$\a\e^{(1-\Theta)\mud}-\a\e^\mud-2\b-\g+\g\e^{\Theta\mud}=0$.
Using~(\ref{abgII}), this condition simplifies to
\beq
\g(\e^{\Theta\mud}-\e^{\mus+\mud})(1-\e^{-\Theta\mud})=2\b.
\label{equals2b}
\eeq
In contrast to Model~I,
we now obtain a transient forgetting exponent
\beq
\Theta=\frac{1}{\mud}\ln\left[\frac{\b}{\g}+\frac{1}{2}\left(\e^{\mus+\mud}+1
+\sqrt{\left({\ss\e^{\mus+\mud}\!+\!1\!+\!\frac{2\b}{\g}}\right)^2\!-4\e^{\mus+\mud}}\,\right)
\right],
\label{Theta}
\eeq
which depends continuously on $\b$.
It turns out to be a strongly increasing function of~$\b$,
starting from the universal value $\theta$ in the $\b\to0$ limit as
\beq
\Theta=\theta+\frac{2\b}{(\e^{\mus+\mud}-1)\mud\g}+\cdots
\eeq
and reaching its maximum for $\b=\b_\max(\g)$ (see~(\ref{bII})).
Figure~\ref{Th} shows a plot of $\Theta$ against $\b$ for the
parameters~(\ref{pars}).

The occurrence of the non-trivial equation~(\ref{equals2b})
for the exponent $\Theta$ in the case of Model~II
can be traced back to the difference in architecture between both models.
For Model~I, where $\b$-transitions involve a non-local reinjection
to the uppermost level,
the rates multiplying $S_n$ and $D_n$ for generic $n$
in the right-hand side of~(\ref{meanI}) are identical and involve the combination
$\a_n+\b_n+\g_n$.
For Model~II, where $\b$-transitions take place locally at any depth,
the rates multiplying $S_n$ and $D_n$ for generic $n$
in the right-hand side of~(\ref{meanII}) are different,
as they are respectively proportional to $\a_n+\g_n$ and $\a_n+2\b_n+\g_n$,
so that {\it locally} the polarisation $D_n$
relaxes more rapidly than the level occupation $S_n$.

Finally, the conclusions of this Appendix regarding the forgetting exponents
hold more generally as soon as the level occupation probabilities in the initial state
fall off exponentially more rapidly than the profile~(\ref{sdstat})
of the default state.

\end{document}